\documentclass[11pt, a4paper, logo, copyright, nonumbering]{deepseek}

\usepackage{booktabs}
\usepackage{subcaption}
\usepackage{listings}
\usepackage{xspace}
\usepackage{enumitem}

\usepackage{relsize}

\usepackage[authoryear, sort&compress, round]{natbib}
\usepackage{dblfloatfix}
\usepackage{ulem}
\usepackage{caption}
\usepackage{dramatist}
\usepackage{pifont}
\usepackage{multirow}
\usepackage{tcolorbox}
\usepackage{xltabular}
\usepackage{diagbox}
\usepackage{longtable}
\usepackage{hyperref}

\usepackage{algorithm}
\usepackage{algpseudocode}
\usepackage{amsfonts}
\usepackage{amsmath}
\usepackage{amssymb}
\usepackage{lineno}
\usepackage{adjustbox}
\usepackage[bottom]{footmisc}

\usepackage{CJKutf8}
\usepackage{setspace}

\usepackage{dsfont}
\usepackage{array}
\usepackage{tabularx}
\usepackage{xcolor}

\usepackage{lipsum}
\usepackage{multicol}

\makeatletter
\def\@BTrule[#1]{%
  \ifx\longtable\undefined
    \let\@BTswitch\@BTnormal
  \else\ifx\hline\LT@hline
    \nobreak
    \let\@BTswitch\@BLTrule
  \else
     \let\@BTswitch\@BTnormal
  \fi\fi
  \global\@thisrulewidth=#1\relax
  \ifnum\@thisruleclass=\tw@\vskip\@aboverulesep\else
  \ifnum\@lastruleclass=\z@\vskip\@aboverulesep\else
  \ifnum\@lastruleclass=\@ne\vskip\doublerulesep\fi\fi\fi
  \@BTswitch}
\makeatother

\addto\extrasenglish{
}

 {\begin{list}{}%
         {\setlength{\leftmargin}{#1}}%
         \item[]%
 }
 {\end{list}}

\reportnumber{001}

\renewcommand{\today}{}
\usepackage{wasysym}
\newcommand{\loo}{\Circle\kern0.15em\Circle\kern0.15em\Circle}
\newcommand{\lo}{\CIRCLE\kern0.15em\Circle\kern0.15em\Circle}
\newcommand{\lom}{\CIRCLE\kern0.15em\LEFTcircle\kern0.15em\Circle}
\newcommand{\md}{\CIRCLE\kern0.15em\CIRCLE\kern0.15em\Circle}
\newcommand{\hi}{\CIRCLE\kern0.15em\CIRCLE\kern0.15em\CIRCLE}

\newcommand{\systemname}{DSec\xspace}
\newcommand{\sdkname}{\cc{libdsec}\xspace}

\setlist{nosep}
\AddToHook{env/table/begin}{}
\renewcommand{\phi}{\varphi}

\newcommand{\cc}[1]{\mbox{\smaller[0.5]\texttt{#1}}}

\renewcommand{\epsilon}{\varepsilon}
\renewcommand{\imath}{\mathrm{i}}

\AtBeginDocument{%
    \def\Snospace~{\S{}}

}

\newlength{\restsubwidth}
\newlength{\restsubheight}
\newlength{\restsubmoreheight}
\newcommand{\rest}[2]{%
        \settowidth{\restsubwidth}{\ensuremath{#2}}
        \settoheight{\restsubheight}{\ensuremath{{}_{#2}}}
        \ensuremath{{#1\hskip 0.5pt}_{\vrule\kern2pt\parbox[b][%
        4pt][b]{\the\restsubwidth}{%
                        \ensuremath{{}_{#2}}}}}
        }

\newcommand{\ProdData}[1]{#1}

\newcommand{\ProdNodesPerUnit}{\ProdData{160}}
\newcommand{\ProdCpuCoresPerUnitK}{\ProdData{30}}
\newcommand{\ProdMemoryPerUnitTB}{\ProdData{250}}

\newcommand{\ProdDailySandboxesM}{\ProdData{3}}
\newcommand{\ProdPeakConcurrentSandboxesK}{\ProdData{380}}
\newcommand{\ProdPeakConcurrentSandboxes}{\ProdData{380,000}}

\newcommand{\ProdCreationRate}{\ProdData{5{,}000}}

\newcommand{\ProdMaxSandboxesPerJobK}{\ProdData{32}}
\newcommand{\ProdMicroVMsPerNode}{\ProdData{800}}
\newcommand{\ProdContainersPerNode}{\ProdData{3{,}200}}

\newcommand{\ProdCloudImageSetTB}{\ProdData{30}}
\newcommand{\ProdCloudTaskCoveragePct}{\ProdData{70}}
\newcommand{\ProdCloudOffloadThresholdPct}{\ProdData{80}}
\newcommand{\ProdCloudVMsPerUnit}{\ProdData{200}}
\newcommand{\ProdCloudOverflowSharePct}{\ProdData{30}}

\newcommand{\ProdWorkloadToolkits}{\ProdData{103}}
\newcommand{\ProdWorkloadExtraLayerPct}{\ProdData{67.8}}
\newcommand{\ProdWorkloadArtifactsLowerBoundTB}{\ProdData{130}}

\newcommand{\ProdStorageSSDsPerServer}{\ProdData{20}}
\newcommand{\ProdStorageSSDCapacityTB}{\ProdData{15}}
\newcommand{\ProdStorageNICsPerServer}{\ProdData{2}}
\newcommand{\ProdStorageServerCount}{\ProdData{Tens of}}
\newcommand{\ProdStorageCpuCoreCount}{\ProdData{hundreds of thousands of}}

\newcommand{\ProdDockerPatchLines}{\ProdData{30}}

\title{\centering DeepSeek Elastic Compute (\systemname): \\ A Sandbox Infrastructure for Effective Agentic Training at Scale}

\author{%
\footnotesize
\begingroup
\spaceskip=0.41em plus 0.06em minus 0.03em
\xspaceskip=\spaceskip
Jialiang Huang$^{\dagger\ddagger}$, Hongxuan Tang$^{\dagger}$, Jingchang Chen$^{\dagger}$, Yuxuan Liu$^{\dagger}$, Yixiao Chen$^{\dagger}$, Yuan Cheng$^{\dagger}$, 
Yi Tao$^{\dagger}$, Jingli Zhou$^{\dagger}$, Yupeng Chen$^{\dagger}$, Haoyu Chen$^{\dagger}$, Jiarui Wang$^{\dagger}$, Shengkai Lin$^{\dagger}$, Chuqi Zhang$^{\dagger}$, Bryan Lee Teng$^{\dagger}$, Lian Guo$^{\dagger}$, 
Zhe Fu, Wenjun Gao, Yisong Wang, Liang Zhao, Zehao Wang, Ziwei Xie, Yongqiang Guo, Peixin Cong, Ziyi Gao, Shuiping Yu, Hanwei Xu, Zuofan Wu, Zhizhou Ren, Yuyang Zhou, Bowei Zhang, Zhihuan Huang, Qihao Zhu, Lei Wang, Tianle Lin, Han Yu, Jiewen Hu, Dejian Yang, Shuo Yang, Shanghao Lu, Shaoyuan Chen, Junjie Qiu, Zhangli Sha, Yinmin Zhong, Yongtong Wu, Shiyu Wang, Wei Liu, Bingzheng Xu, Longhao Chen, Qiushi Du, Yuzhen Huang, Shirong Ma, Yaohui Wang, Mingshu Chen, Tongrui Xiong, Y.C. Yan, Haowen Luo, Haofen Liang, Xiaokang Zhang, Weihao Zeng, Runxin Xu, Peiyi Wang, Jinhua Zhu, Ruoyu Zhang, Wenkai Yang, Qi Tang, Jiping Yu, Tian Ye, Ruizhe Pan, Honghui Ding, Xiaodong Liu, Lingxiao Luo, Zhihong Shao, Yuhan Wu, Jibai Lu, Wen Liu, Haoling Zhang, Jingcheng Hu, Yaoyang Ye, Chaofan Lin, Zhaochen Zhang, Jianan Tong, Hengxu Wu, Zhihao Li, Yicheng Wang, Luyao Wang, Yuzhuo Bai, Lingyue Fu, Ruifan Xu, Y.Z. Wang, Zonglin Li, Mingqi Wei, Haiyang Shen, Chengyuan Zhang, Chao Jin, Zili Zhang, R.H. Yang, Xinbo Xu, Jian Zhou, Ruidong Zhu, Yuzhe Guo, Zelun Pan, Shaoheng Nie, Erhang Li, Shuhan Lin, Zheng Liu, Anshuo Chen, Zilong Lyu, Sinuo Cao, Rui Yu, Chuhao Wang, Junyi Guo, Junxiao Song, Kaifeng Chen, Menghao Ye, Junxian Li, Di Wu, Haiyang Ma, Yilun Wang, Haoran Yang, Yizai Cai, Shichun Liu, Yiping Wang, Junbo Sun, Shicheng Xu, 
Xiao Bi, Ying He, Yichao Zhang, 
Mingxing Zhang$^{\ddagger}$, Liyue Zhang$^{*\dagger}$, Panpan Huang, Wenfeng Liang
\endgroup\\
{\small DeepSeek-AI \quad $^{\ddagger}$Tsinghua University \\
\ttfamily\small research@deepseek.com}
}
\correspondingauthor{%
\footnotesize
$^{*}$Corresponding author.\ \ $^{\dagger}$DSec project developers.\ \ $^{\ddagger}$Tsinghua University.\\
Jialiang Huang is a Ph.D.\ student advised by Mingxing Zhang.
He contributed to this work during an internship at DeepSeek-AI under the mentorship of Liyue Zhang.%
}

\hypersetup{
  pdfauthor={Jialiang Huang, Hongxuan Tang, Jingchang Chen, Yuxuan Liu, Yixiao Chen, Yuan Cheng, Yi Tao, Jingli Zhou, Yupeng Chen, Haoyu Chen, Jiarui Wang, Shengkai Lin, Chuqi Zhang, Bryan Lee Teng, Lian Guo, Zhe Fu, Wenjun Gao, Yisong Wang, Liang Zhao, Zehao Wang, Ziwei Xie, Yongqiang Guo, Peixin Cong, Ziyi Gao, Shuiping Yu, Hanwei Xu, Zuofan Wu, Zhizhou Ren, Yuyang Zhou, Bowei Zhang, Zhihuan Huang, Qihao Zhu, Lei Wang, Tianle Lin, Han Yu, Jiewen Hu, Dejian Yang, Shuo Yang, Shanghao Lu, Shaoyuan Chen, Junjie Qiu, Zhangli Sha, Yinmin Zhong, Yongtong Wu, Shiyu Wang, Wei Liu, Bingzheng Xu, Longhao Chen, Qiushi Du, Yuzhen Huang, Shirong Ma, Yaohui Wang, Mingshu Chen, Tongrui Xiong, Y.C. Yan, Haowen Luo, Haofen Liang, Xiaokang Zhang, Weihao Zeng, Runxin Xu, Peiyi Wang, Jinhua Zhu, Ruoyu Zhang, Wenkai Yang, Qi Tang, Jiping Yu, Tian Ye, Ruizhe Pan, Honghui Ding, Xiaodong Liu, Lingxiao Luo, Zhihong Shao, Yuhan Wu, Jibai Lu, Wen Liu, Haoling Zhang, Jingcheng Hu, Yaoyang Ye, Chaofan Lin, Zhaochen Zhang, Jianan Tong, Hengxu Wu, Zhihao Li, Yicheng Wang, Luyao Wang, Yuzhuo Bai, Lingyue Fu, Ruifan Xu, Y.Z. Wang, Zonglin Li, Mingqi Wei, Haiyang Shen, Chengyuan Zhang, Chao Jin, Zili Zhang, R.H. Yang, Xinbo Xu, Jian Zhou, Ruidong Zhu, Yuzhe Guo, Zelun Pan, Shaoheng Nie, Erhang Li, Shuhan Lin, Zheng Liu, Anshuo Chen, Zilong Lyu, Sinuo Cao, Rui Yu, Chuhao Wang, Junyi Guo, Junxiao Song, Kaifeng Chen, Menghao Ye, Junxian Li, Di Wu, Haiyang Ma, Yilun Wang, Haoran Yang, Yizai Cai, Shichun Liu, Yiping Wang, Junbo Sun, Xiao Bi, Ying He, Yichao Zhang, Mingxing Zhang, Liyue Zhang, Panpan Huang, Wenfeng Liang
}
}

\begin{abstract}
Large-scale agentic training and evaluation with large language models (LLMs) rely on isolated, stateful execution environments in which models inspect repositories, invoke tools, execute commands, and interact with task-specific services.
These workloads create sandboxes in large bursts, span heterogeneous functionality and isolation requirements, retain state across long interactions, and draw from large image corpora with limited reuse.
Supporting them therefore requires an elastic execution platform rather than a single sandbox runtime.

This report presents DeepSeek Elastic Compute (\systemname{}), a production sandbox platform that exposes FnCall, container, microVM, and full-VM sandbox backends through a unified SDK.
\systemname{} coordinates placement and lifecycle management across the cluster, composes environments from independently versioned layers, combines memory sharing, reclamation, and CPU scheduling for high-density execution, and loads image data on demand from Fire-Flyer File System (3FS), a cluster-wide distributed filesystem.
\systemname{} is co-designed with the reinforcement learning (RL) framework, decouples stateful rollout execution from preemptible GPU training, coordinates sandbox lifecycle with training to preserve rollout state while reclaiming idle resources, and mitigates agent misbehavior such as reward hacking.

A single production-scale unit of \systemname{} spans around \ProdNodesPerUnit{} nodes, serving about \ProdDailySandboxesM{} million sandboxes per day; in production, it supports over \ProdPeakConcurrentSandboxes{} concurrent sandboxes and sustains over \ProdCreationRate{} sandbox creations per second.
Our evaluation and deployment experience show that these mechanisms reduce environment setup and image-distribution overhead, improve memory efficiency, and preserve latency-sensitive performance under high-density overcommit.
\end{abstract}

\begin{document}
\maketitle

\fancyhead{}
\begin{CJK*}{UTF8}{gbsn}

\section{Introduction}
\label{sec:intro}

Recent advances in frontier LLMs have made agentic workflows practical and widely adopted \citep{deepseek-r1,gpt4,swe-bench}.
Instead of producing a single text answer, an agentic model interacts with an execution environment: it may navigate codebases, call tools, execute commands, inspect failures, and modify files, or operate browsers and desktop applications through graphical interfaces in computer-use tasks~\citep{xie2024osworld,zhou2024webarena}.
Across these workloads, the model iterates based on feedback until a task is solved.
This execution model has led to a growing ecosystem of agent tools and orchestration harnesses, such as DeepSeek Harness (DSH)~\citep{shi2026programmingparadigmspatiotemporalcomposability}, OpenCode~\citep{opencode}, and multi-agent training harnesses.
Training reliable agents requires reinforcement learning (RL) at scale, in which models learn through interaction with real, isolated execution environments rather than solely from static input-output examples.

The agentic training pipeline encompasses environment and data construction, RL rollouts, reward computation, policy updates, and periodic evaluation.
Among these stages, RL rollout and evaluation impose the highest pressure on the sandbox platform because they are large-scale, concurrent, and tightly coupled with the training loop.
In RL~\citep{rlhf,deepseek-r1}, training proceeds as a feedback loop with three stages.
First, during rollout, the current model interacts with the sandboxed environment: it reads files, issues tool calls, executes commands, observes outputs, and produces a trajectory for each task.
Second, during reward computation, the framework scores the trajectory using native execution signals such as exit codes, stdout, test pass rates, or task-specific verifiers.
Third, during policy update, the RL algorithm updates the model parameters from the collected trajectories and rewards.
Periodic evaluation follows a similar execution path, except that the resulting trajectories are used to measure model capability rather than to update parameters.
Recent systems further pipeline generation and policy optimization through asynchronous rollouts, continuously replenishing completed samples to maintain high concurrency and mitigate long-tail stragglers~\citep{deepseek2026v41}.
For agentic workloads, this design keeps many stateful sandbox sessions in flight and may interrupt and resume their associated rollouts across policy updates or scheduler preemptions, further increasing the platform's concurrency, lifecycle-management, and state-consistency requirements.

For each rollout or evaluation task, the platform must materialize an isolated task-specific environment, including its repositories, dependencies, services, evaluation scripts, and coding harnesses.
The environment must be close enough to a real machine to run unmodified software stacks, package managers, build tools, browsers, emulators, and task-specific services.
A robust, high-throughput sandbox runtime is therefore foundational for obtaining accurate and verifiable RL and evaluation results.

Agentic sandbox workloads have several properties that shape the platform design:

\begin{enumerate}[
    label=(\arabic*),
    wide=2em,
    labelsep=0.5em,
    nosep
]
\item \textbf{Rollout and evaluation jobs create sandboxes in a bursty manner.}
A single job may request up to \ProdMaxSandboxesPerJobK{}K sandbox instances, so the platform must accept and place many sandboxes concurrently.
Such bursts make horizontal scalability a system-wide requirement and require shared services, such as scheduling and image distribution, to avoid centralized bottlenecks.
\item \textbf{Sandboxes must run at high density.}
During agent interaction, a sandbox often waits for the LLM to generate the next action, so CPU usage is sparse and naturally suitable for overcommit.
For instance, in production, this allows a single node to host up to \ProdMicroVMsPerNode{} microVMs or \ProdContainersPerNode{} containers, but only if the platform can safely overcommit resources and manage lifecycle pressure at node scale.
\item \textbf{Agent sandboxes are stateful and long-lived.}
The model may modify files, install dependencies, and start services, and later tool calls depend on this accumulated state.
Since a sandbox can stay alive across many LLM interaction turns, memory footprint, guest page cache, host page cache, and writable state may remain pinned long after the CPU becomes idle.
Under high-density overcommit, these resident costs directly limit cluster capacity, so memory sharing and reclamation become important platform requirements.

\item \textbf{Agent workloads are highly heterogeneous.}
The platform must cover OJ-like script execution, software-engineering tasks over full repositories, security tasks, computer-use workloads, mobile development environments (e.g., Android), and other full-system environments.
These workloads differ substantially in CPU and memory demand, dependency footprint, required system functionality, and isolation strength.
A single sandbox abstraction cannot cover all of them efficiently.
For example, lightweight function calls are preferable for short stateless tasks, whereas virtual machines (VMs) are better suited to workloads that require a complete commercial off-the-shelf operating system.

\item \textbf{Environment diversity is high even within the same workload class.}
Training and evaluation corpora contain many tasks, and each task may require its own repository, dependency versions, services, toolkits, evaluation scripts, or VM snapshots.
As a result, the platform must serve a large number of distinct images and environment artifacts, with limited reuse for many of them.
Under bursty startup, fetching these diverse task images from a registry would concentrate load on the distribution path, inflate startup latency, and introduce extra I/O that interferes with already-running sandboxes.
In our ablation, eager image pulling stretches completion time by 1.7$\times$, while on-demand loading reduces cumulative disk writes by 57\%.
\item \textbf{Agent execution is untrustworthy.}
Agents may corrupt filesystems, exhaust resources, or interfere with system components, potentially disrupting rollouts or other co-located workloads.
The platform therefore requires fine-grained access control and misbehavior analysis to contain and diagnose agent-induced failures.

\item \textbf{Agent execution is interruptible.}
GPU training jobs may be preempted while long-running rollouts are still in progress.
The platform must therefore preserve execution state and support efficient recovery across interruptions.
\end{enumerate}

These properties define the role of an agent sandbox platform.
DSec provides elastic service scaling, high-density resource management, memory sharing and reclamation, multiple isolation mechanisms for different workload classes, scalable image distribution, and explicit integration with the training framework for preemption-safe resumption, task-specific network policy, and agent misbehaving mitigation.

The rest of this report presents \systemname{} from platform abstraction to implementation and evaluation.
\autoref{sec:overview} introduces \systemname{} from the user perspective, including supported workloads, sandbox backends, and operating scale.
\autoref{sec:architecture} describes the end-to-end platform architecture.
\autoref{sec:challenges} characterizes the production workload and the platform challenges it creates.
\autoref{sec:design} presents the core system mechanisms for environment composition, image distribution, and high-density resource management.
\autoref{sec:codesign} describes co-design with the RL framework for environment construction, state preservation, resource reclamation across preemption, and the analysis of agent misbehavior with targeted access-control mitigations.
\autoref{sec:implementation} summarizes additional implementation details.
\autoref{sec:eval} evaluates the effectiveness of the design, and \autoref{sec:related} discusses related works.

\section{Overview of \systemname}
\label{sec:overview}

This chapter presents the user-facing view of \systemname{}.
From the platform's perspective, users are the training frameworks, evaluation frameworks, and data-construction pipelines that call the software development kit (SDK) on behalf of researchers; we refer to them collectively as users throughout the report.
It covers the SDK entry point, the sandbox backends exposed by the platform and the workload classes they serve, the lifecycle of a sandbox session, and the operating scale of the production deployment.

\subsection{SDK Entry Point}
\label{subsec:overview-sdk}

Users access \systemname{} through \sdkname{}, a Python client library for the sandbox service.
\sdkname{} gives users a unified SDK entry point for creating and operating sandboxes, while still requiring them to choose the sandbox backend appropriate for the task.
A typical request specifies the sandbox type, image or environment identifier, CPU and memory limits, lifetime settings, network rules, and initial user context.
After creation, the user can execute shell commands or tool calls and collect command outputs and return status.
\autoref{lst:sdk} shows a minimal container session: the client connects to the service endpoint, requests a sandbox with the desired resource and network policy, runs a command, and releases it.

\begin{lstlisting}[language=Python,caption={A minimal sandbox session through \sdkname{}.},label={lst:sdk},basicstyle=\ttfamily\scriptsize]
client = DSecClient()
await client.open()
args = DSecContainerRunArgs(
    container_image="registry.../sphinx-9658:official",
    memory_limit_mb=4096, cpu_cores_limit=4,
    ttl_running_stop=300,      # idle timeout
    network_rules={"npm": False, "pypi": True},
    init_user="root",
)
sandbox = await client.run_container(args, timeout=120)
result = await sandbox.run_shell("echo hello world")
await sandbox.stop()
\end{lstlisting}

In this example, the network rules allow access to PyPI (\texttt{pypi=True}) but deny access to NPM (\texttt{npm=False}).
This fine-grained network control is discussed in detail in \autoref{sec:codesign}.
This interface is intentionally not a full semantic abstraction over all backends.
Function calls, containers, microVMs, and full VMs have different startup costs, isolation boundaries, filesystem semantics, and operating-system capabilities.
\sdkname{} provides a unified access path and a similar operational model, but the caller remains responsible for selecting a backend that matches the workload.

\subsection{Sandbox Backends}
\label{subsec:overview-backends}

Sandbox runtimes face a fundamental tension: stronger isolation and more complete system functionality usually come with higher startup latency and resource overhead.
Since no single sandbox abstraction fits all agentic tasks, \systemname{} supports multiple backends spanning this tradeoff space.
\autoref{tab:overview-workload-fit} summarizes their typical fit.

\begin{table}[t]
\centering
\caption{Typical workload fit across \systemname{} sandbox backends.}
\label{tab:overview-workload-fit}
\small
\setlength{\tabcolsep}{4pt}
\begin{tabular*}{\linewidth}{@{\hspace{6pt}\extracolsep{\fill}}lcccc@{\hspace{6pt}}}
\toprule
\textbf{Characteristic} & \textbf{FnCall} & \textbf{Container} & \textbf{MicroVM} & \textbf{Full VM} \\
\midrule
Runtime Performance & \hi & \md & \lom & \lo \\
Dependency footprint & \loo & \hi & \hi & \md \\
Isolation level & \loo & \md & \hi & \hi \\
Full OS functionality & \loo & \lo & \md & \hi \\
Resource overhead & \loo & \lo & \md & \hi \\
\midrule
Scenarios  & OJ-like tasks & SWE & Security & COTS OS \\
           & GPU kernel exec & Tool use & Computer use & Graphics \\
\bottomrule
\end{tabular*}
\vspace{2pt}

More \CIRCLE = higher demand.
\end{table}

\textbf{FnCall} targets short, stateless tasks such as OJ workloads, code compilation, serverless programs, GPU kernels, and utility code.
FnCall tasks run in reusable precreated CPU or GPU containers, avoiding per-invocation provisioning overhead.
For GPU workloads, FnCall supports (i) \textit{shared} mode, where multiple containers share a GPU instance, maximizing utilization for lightweight workloads, and (ii) \textit{exclusive} mode, where one container reserves a GPU instance during its lifecycle for performance-sensitive tasks (e.g., operator evaluation).
\textbf{Containers} are the main backend for software-engineering and general tool-use workloads.
They provide fast startup and high packing density, and they run the Linux software stacks used by most repository-level tasks.
Their main limitation is that they share the host kernel, which is not always appropriate for security-sensitive tasks.
\textbf{Firecracker microVMs}~\citep{firecracker} provide a stronger isolation boundary while retaining Linux compatibility.
They are useful for security-sensitive tasks, stronger tenant isolation, and workloads that need a VM boundary with Linux compatibility.
This comes at higher memory overhead and slower startup than containers.
\textbf{Full VM backends} cover workloads that require a complete commercial off-the-shelf operating system environment, such as Android VMs through QEMU~\citep{qemu}, as well as those that require a GUI or graphics rendering.
These backends have the highest resource overhead, but they are necessary for tasks that depend on OS-specific APIs, mobile runtime behavior, or full-system execution.

In production, containers and microVMs dominate both instance count and resource consumption.
FnCall serves a large number of lightweight invocations with a small set of resident environments, while full VM backends cover specialized but important workload classes.

\subsection{User-Visible Lifecycle}
\label{subsec:overview-lifecycle}

Although the supported backends differ internally, users see a unified high-level lifecycle.
First, the caller creates a sandbox by selecting a backend and specifying the environment artifact, resource limits, lifetime policy, and network policy.
The environment artifact varies by backend and workload.
For containers and microVMs, it is a base image together with task-specific workspace and toolkit layers, which the platform composes into the running environment.
For full VM workloads, it is a prepared VM image or snapshot.
For FnCall, it is a task specification containing the task type, dependency files, and the code or script to run.
These artifacts become the basis for the environment composition and image-distribution mechanisms discussed later in the report.
Second, the platform prepares the environment and makes it ready for interaction.
Third, the user issues commands or tool calls, observes outputs, and runs task-specific checks or tests.
A sandbox is stateful throughout its lifetime: file edits, installed dependencies, and started services persist across calls, so later commands observe the effects of earlier ones.
Because a sandbox stays alive across many interaction turns while its CPU is often idle between them, its resident state remains pinned long after the last command, one of the high-density challenges characterized in~\autoref{sec:challenges}.
Finally, the sandbox is stopped explicitly or reclaimed once its time-to-live elapses, so that idle or abandoned sessions do not hold resources indefinitely.

\subsection{Deployment Scale}
\label{subsec:overview-scale}

\systemname{} is deployed across multiple scale units that share a 3FS~\citep{hf3fs_repo} distributed file system deployment for base images and workspace storage.
Within one scale unit, the platform spans nearly \ProdNodesPerUnit{} CPU nodes with \ProdCpuCoresPerUnitK{}K cores and $\sim$\ProdMemoryPerUnitTB{}\,TB of DRAM.
It manages petabytes of layers and images.
On a typical day, a single scale unit serves about \ProdDailySandboxesM{}\,M sandbox instances, with peak concurrency reaching $\sim$\ProdPeakConcurrentSandboxesK{}K and a creation rate exceeding \ProdCreationRate{} instances per second.

These numbers are important for understanding the rest of the report.
\systemname{} is not a single sandbox runtime or a thin wrapper around containers.
It is a production execution platform that must combine user-facing sandbox abstractions, backend-specific runtimes, scalable image storage, high-density resource management, and training-framework integration.

\section{Platform Architecture}
\label{sec:architecture}

\autoref{sec:overview} presented \systemname{} as users see it: an SDK, a set of sandbox backends, and a session lifecycle.
This chapter turns to the platform behind that interface and describes how a request travels from the SDK to a running sandbox and which components it passes through.
We describe the architecture in terms of cluster-level services and the sandbox runtime.
Cluster-level services provide request ingress, identity and access management, sandbox placement, and a view of cluster health and load.
The sandbox runtime handles node-local admission, sandbox creation, execution, and resource reclamation, relying on 3FS for image data.

\subsection{Overview}
\label{subsec:arch-overview}

\begin{figure}[!ht]
    \centering
    \includegraphics[width=1.0\linewidth]{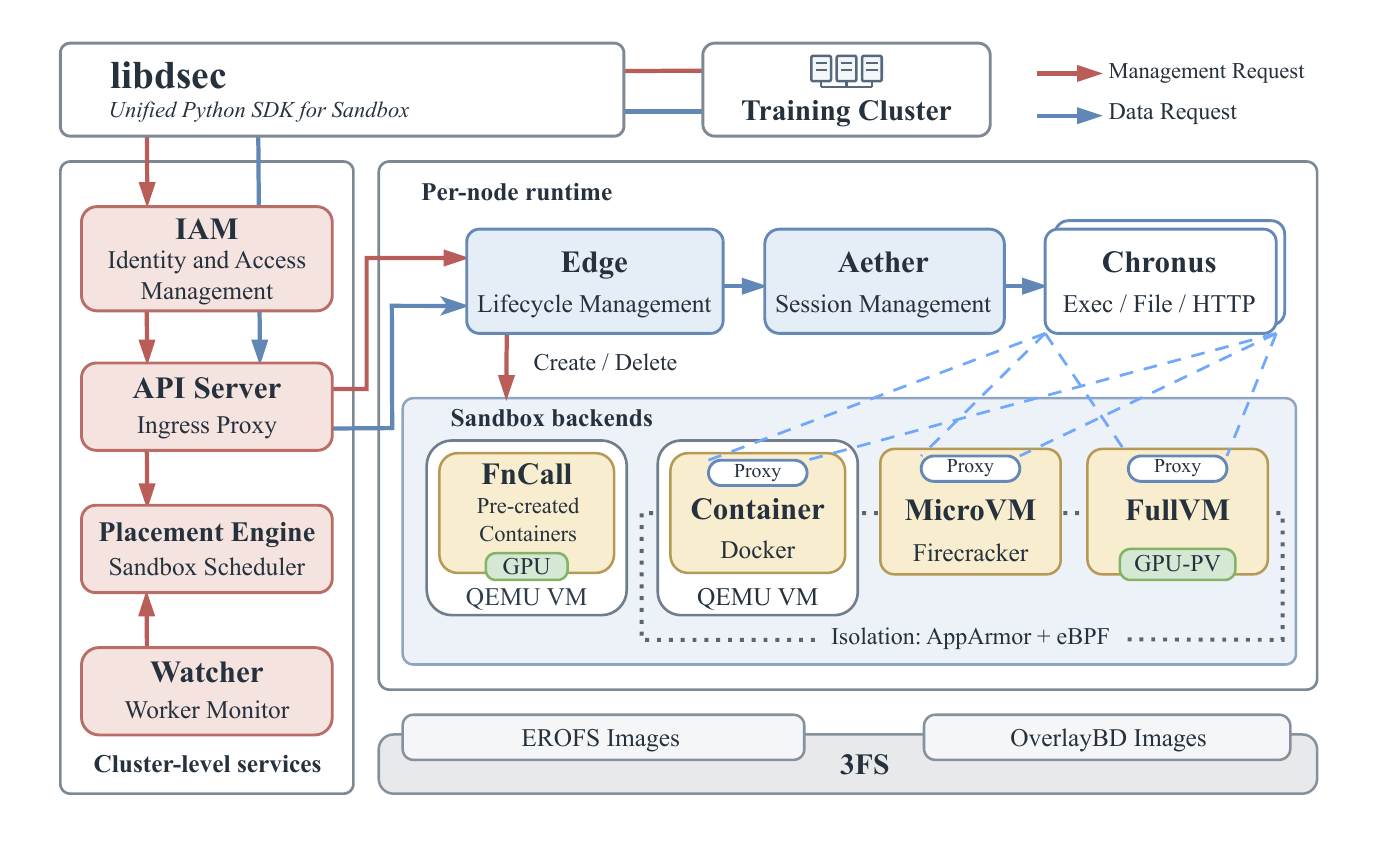}
    \caption{\systemname{} architecture. Each proxy mediates communication between a container or VM sandbox and the rest of the platform. FnCall follows a separate execution path and does not use this proxy.}
    \label{fig:system_arch}
\end{figure}

At a high level, a sandbox creation request is first sent to \cc{IAM} for authentication and authorization.
Once authorized, the request proceeds to the \cc{placement engine}, which selects a target node using health and load information collected by the \cc{watcher}.
After placement, the \cc{apiserver} forwards the request to the \cc{edge} on that node.
The \cc{edge} then checks local capacity, creating the sandbox with the requested backend if capacity permits and rejecting the request otherwise.
Image data needed by the sandbox is stored in 3FS and fetched on demand during startup and execution.
Container, microVM, and full VM sandboxes run a per-sandbox proxy (\cc{aether}) and one or more \cc{chronus} instances for command execution, filesystem access, and other runtime operations.
After one of these sandboxes is running, its operations are routed through the \cc{apiserver}, \cc{edge}, \cc{aether}, and \cc{chronus}.
FnCall, by contrast, uses neither \cc{aether} nor \cc{chronus} and follows a separate request path: the submitted task is executed directly in a precreated container, followed by best-effort cleanup of task state.

\subsection{Cluster-Level Services}
\label{subsec:arch-services}

Cluster-level services manage access to the platform and coordinate sandbox requests across compute nodes.
They comprise \cc{IAM}, the \cc{apiserver}, the \cc{placement engine}, and the \cc{watcher}.

\noindent \textbf{IAM.}
Identity and Access Management (\cc{IAM}) authenticates callers and authorizes all management requests to \systemname{}.
For example, requests to create or delete sandboxes or change a user's resource or concurrency limits must pass \cc{IAM} checks before execution.
A principal is the user or service identity associated with a management request.
\cc{IAM} uses projects to define scopes for resource management and access control.
Within a project, access policies specify which principals may perform which management operations on its resources, while resource quotas limit resource consumption.

We support multi-level project nesting rather than the flat or two-level hierarchies common in cloud platforms.
Authorized principals, including agents and harnesses, can create subprojects, delegate part of the parent quota, and grant management permissions within them.
Delegation is bounded by the parent: a principal cannot grant permissions it does not hold, and subproject policies and quotas cannot exceed the parent's access-control or resource limits.
Humans and agents use the same management API and authorization model.

\noindent \textbf{API Server.}
The \cc{apiserver} serves as the ingress proxy for the sandbox cluster.
Training and evaluation code invokes \sdkname{} from trusted GPU servers, while sandboxes execute untrusted model-generated code and may access external networks.
The two sides are therefore network-isolated, with the \cc{apiserver} as the only permitted communication path.
All sandbox requests, including creation, command execution, and streaming I/O, pass through this ingress.
The \cc{apiserver} maintains no per-sandbox state.
It periodically refreshes the set of \cc{edge} nodes from the \cc{watcher}, while each sandbox ID encodes its owning \cc{edge}.
Any \cc{apiserver} instance can therefore resolve and forward a request directly to the target \cc{edge}, enabling the ingress tier to scale horizontally.

\noindent \textbf{Placement Engine.}
The \cc{placement engine} selects a host node for each new sandbox.
Placement proceeds in two stages: filtering and ranking.
The filtering stage retains only healthy nodes that provide the backend and hardware capabilities required by the request.
For example, a request for a GPU-enabled sandbox is restricted to nodes equipped with the required GPUs.
The ranking stage randomly samples a few eligible nodes and selects the least loaded among them.

\noindent \textbf{Watcher.}
The \cc{placement engine}'s decisions are only as good as its view of the fleet, which the \cc{watcher} provides.
The \cc{watcher} periodically probes the health of each \cc{edge} and host and collects scheduling-relevant state, such as the number of running sandboxes across backend types, broken down per \cc{edge}, per user, and per task.
The \cc{placement engine} periodically pulls this state from the \cc{watcher} and uses the latest view when evaluating new creation requests.

Please note that neither the \cc{placement engine} nor the \cc{watcher} requires durable state.
The \cc{placement engine} keeps no sandbox execution state, and the \cc{watcher} can rebuild its fleet view after a restart by polling the \cc{edges} again.
This makes \cc{placement engine} and \cc{watcher} instances easy to add or replace without a costly recovery step.

\subsection{Sandbox Runtime}
\label{subsec:arch-runtime}

The sandbox runtime creates and operates individual sandboxes and manages their resources.
It includes \cc{edge}, \cc{aether}, and \cc{chronus}, and relies on 3FS for shared image storage.

\noindent \textbf{Edge.}
Each node runs an \cc{edge}, a per-machine component that handles creation requests from the \cc{apiserver} for container, microVM, QEMU-based full VM, and FnCall backends.
Before accepting a creation request, the \cc{edge} checks the node's current capacity and rejects the request if capacity is insufficient.
This node-local admission check complements the \cc{placement engine}'s placement decision, which is based on periodically refreshed cluster state.
During creation, the \cc{edge} provisions storage, applies the eBPF-based network policy, and launches the runtime.

FnCall and containers run inside QEMU/libvirt VMs rather than directly on the host.
The VM provides an isolated kernel and network stack and serves as an additional security boundary between untrusted containers and the bare metal.
To better support graphics-intensive workloads, such as computer-use GUI applications, browsers, video games, and 3D rendering, we leverage para-virtualized GPU interfaces of the host hypervisor (e.g., virtio-gpu).
Within the full VM, we support both workloads whose graphics APIs are natively compatible with the host OS, as well as those whose rendering stacks can be translated into host-native APIs through compatibility layers such as DXVK \citep{dxvk}.

Besides tracking the sandbox lifecycle, \cc{edge} coordinates disk and memory snapshots and releases node-local resources when the sandbox stops or its TTL expires.

\noindent \textbf{Aether.}
Container and VM sandboxes run \cc{aether}, a cross-platform proxy that establishes a communication channel with the \cc{edge}.
This \cc{edge}-to-\cc{aether} channel uses a platform-specific transport, such as a Unix domain socket for Linux containers or vsock for VM backends.
The \cc{edge} monitors sandbox health through the channel and marks the sandbox as failed if the channel closes.
For each operation, \cc{aether} uses the operation's terminal-session identifier to create or locate the corresponding \cc{chronus} instance, then forwards the operation over the local channel.
When the session ends, \cc{aether} terminates the corresponding \cc{chronus} process tree.

\noindent \textbf{Chronus.}
\cc{chronus} provides a shell-session abstraction inside the sandbox, with each instance representing one independent shell session.
It exposes cross-platform interfaces for command execution, filesystem operations, HTTP requests, and streaming I/O.
Multiple \cc{chronus} instances can run concurrently within the same sandbox.
Together, \cc{aether} and \cc{chronus} allow \sdkname{} to expose a unified interface for container and VM sandbox operations.

\noindent \textbf{Base image and workspace storage.}
The sandbox runtime uses 3FS as a shared backing store for base images and workspace images.
Container images are converted offline from OCI into EROFS, which separates metadata from data so that metadata is kept local while image data remains in 3FS.
MicroVM disk images use an OverlayBD~\citep{dadi} format over the same storage.
Together, these image formats support on-demand loading and incremental snapshots over a shared base, allowing an \cc{edge} to start a sandbox without first pulling a full image.
\autoref{sec:challenges} characterizes the workload pressures that make scalable image distribution necessary, while \autoref{subsec:image-diversity} describes the corresponding on-demand loading mechanism.

\subsection{Cloud Bursting with Selective Offloading}
\label{subsec:arch-cloud-offload}
\systemname{} uses cloud VMs to absorb transient peaks in sandbox demand while serving the steady-state workload on-premise.
When on-premise utilization exceeds \ProdCloudOffloadThresholdPct{}\%, the \cc{placement engine} offloads a portion of eligible incoming sandbox creation requests to cloud VMs.

Instead of combining a managed container service with object storage, we reuse the on-premise container runtime and EROFS-based image-loading path on cloud VMs.
The EROFS images reside in a cloud-hosted distributed filesystem and are mounted by the cloud VMs.

Production file-access traces show that a compact, de-duplicated EROFS image set totaling \ProdCloudImageSetTB{}\,TB covers the image files accessed by \ProdCloudTaskCoveragePct{}\% of container tasks.
We synchronize this shared image set to the cloud filesystem offline.
Container tasks whose image dependencies are fully contained in this set are classified as cloud-eligible.
Other tasks remain on-premise.
In production, \ProdCloudVMsPerUnit{} cloud VMs in one scale unit absorb $\sim$\ProdCloudOverflowSharePct{}\% of peak overflow, increasing capacity without over-provisioning the on-premise cluster.

\section{Production Sandbox Workloads and Platform Challenges}
\label{sec:challenges}

This chapter characterizes the production workloads served by \systemname{} and the platform challenges that follow from them.
We report \textit{measurements only for containers and microVMs}, which together account for most sandbox instances and resource consumption in production.
The measurements show a combination that is unusual for conventional execution services: requests arrive in large bursts, each sandbox retains state across an extended interaction, CPU demand is sparse even when many sandboxes are live, and the environment working set is too diverse for node-local image caches.
We connect each workload property to its system consequence here and defer the corresponding mechanisms to the following chapters.

\subsection{Lifecycle and Bursty Demand}
\label{subsec:workload-lifecycle}

\begin{figure}[htbp]
    \centering
    \includegraphics[width=0.7\linewidth]{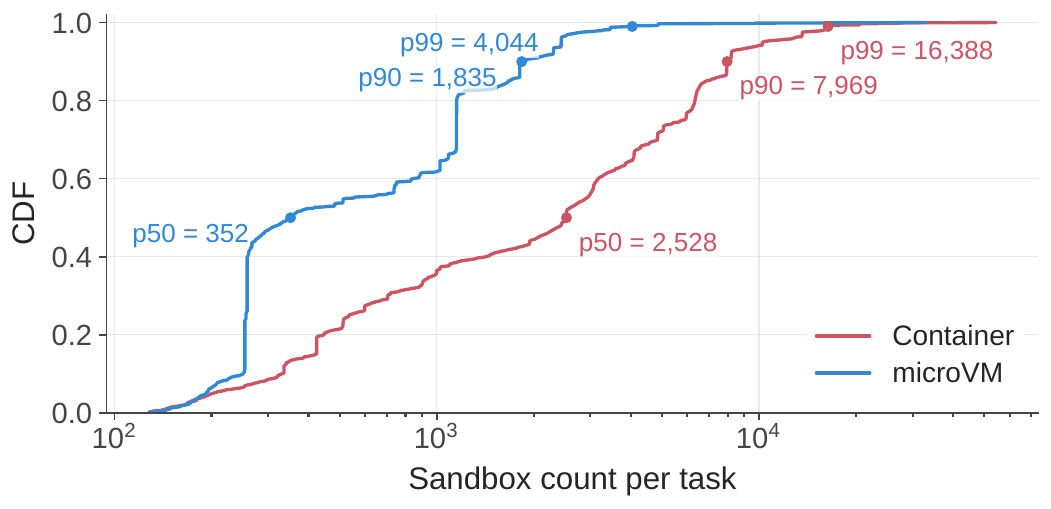}
    \caption{The distribution of the number of sandboxes created per task. Data were sampled over one week in early 2026.}
    \label{fig:sandbox_num_per_task_cdf}
\end{figure}

Rollout and evaluation tasks create sandboxes in batches rather than at a steady rate.
\autoref{fig:sandbox_num_per_task_cdf} shows that a typical container task already creates thousands of sandboxes, and the tail reaches tens of thousands.
The largest production jobs can request up to \ProdMaxSandboxesPerJobK{}K sandboxes.
These requests arrive within a short window because the training or evaluation batch cannot use an instance until its environment is ready.
Consequently, placement, sandbox creation, and environment setup must absorb sharp bursts, while stragglers in any stage delay useful model interaction.

\begin{figure}[htbp]
    \centering
    \includegraphics[width=0.7\linewidth]{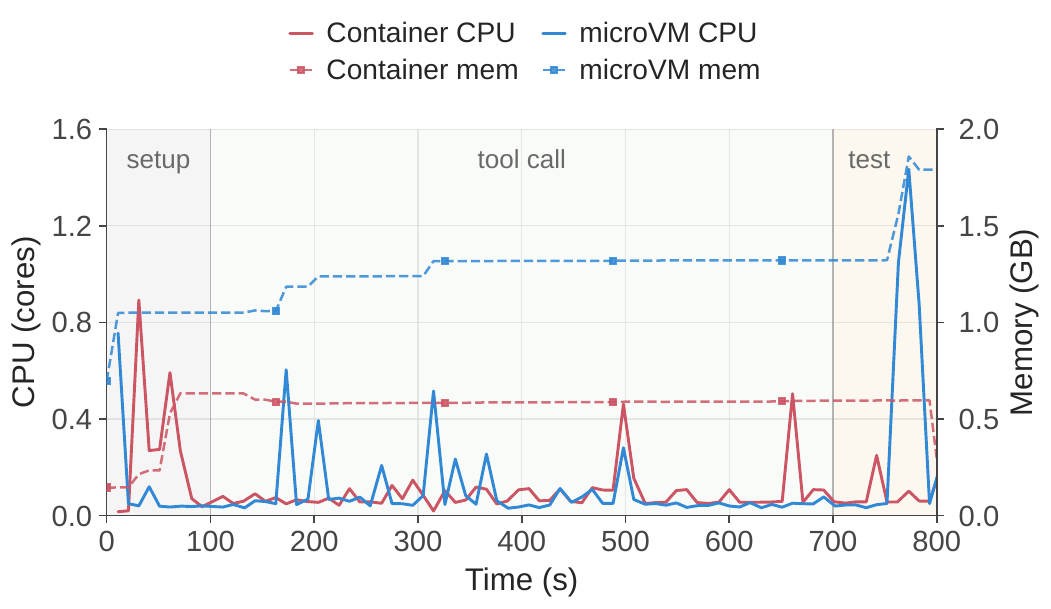}
    \caption{A representative sandbox execution with setup, tool-call, and test phases. CPU demand is intermittent after setup, while the memory footprint and accumulated state persist.}
    \label{fig:sandbox_phase}
\end{figure}

Once created, a sandbox proceeds through three broad phases, as illustrated in \autoref{fig:sandbox_phase}.
The {setup} phase prepares task dependencies, tools, and initialization state.
During the {tool-call} phase, the model alternates between output generation and sandbox operations, producing short CPU bursts separated by periods in which the sandbox waits for the next action.
The {test} phase verifies the result and can briefly increase resource demand again.
These phases do not have fixed durations, but their different resource profiles are important: setup cost is multiplied by burst size, whereas later phases retain sandbox state despite intermittent CPU activity.

\subsection{Environment Diversity and Setup Pressure}
\label{subsec:environment-composition}

The first challenge is a costly setup phase driven by composing each sandbox from independently evolving software components.
A sandbox's content can be decomposed into three parts: a \textit{base image} providing OS-level dependencies (e.g., Ubuntu, Python~3.10, or a Java~8 environment), a \textit{workspace} carrying the task's code repository and its task-specific dependencies, and one or more frequently updated \textit{toolkits} (e.g., the DeepSeek Harness).
During one production week, the container backend served 11,266 base images and 102,171 workspaces, while the microVM backend used two shared base images and 53,590 task-specific workspaces.
The platform also served \ProdWorkloadToolkits{} toolkits, and \ProdWorkloadExtraLayerPct{}\% of sandboxes required at least one workspace or toolkit in addition to the base image.

\begin{table}[ht]
\centering
\caption{Environment artifacts active during one production week in early 2026. MicroVM workspaces are stored as task-specific disk images.}
\small
\setlength{\tabcolsep}{6pt}
\begin{tabular}{lrrrr}
\toprule
\textbf{Backend} & \textbf{Base images} & \textbf{Workspaces} & \textbf{Snapshots} & \textbf{Aggregate size} \\ \midrule
Container & 11,266 & 102,171 & -- & 82.8 TB \\
MicroVM   & 2 & 53,590 & 4,889 & 50.9 TB \\ \bottomrule
\end{tabular}
\label{tab:image_stat}
\end{table}

\begin{figure}[ht]
    \centering
    \captionsetup[subfigure]{justification=centering,singlelinecheck=true}
    \begin{subfigure}[t]{0.52\linewidth}
        \centering
        \includegraphics[width=\linewidth]{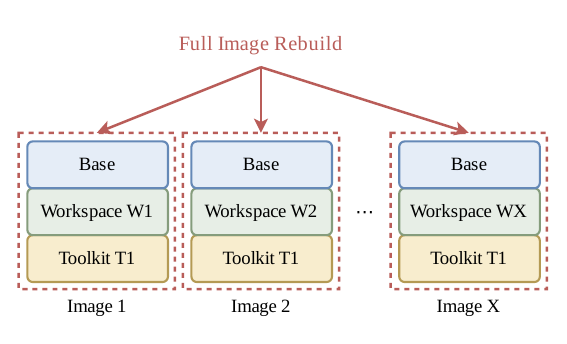}
        \caption{Monolithic images.}
        \label{fig:toolkit-update-monolithic}
    \end{subfigure}
    \hfill
    \begin{subfigure}[t]{0.47\linewidth}
        \centering
        \includegraphics[width=\linewidth]{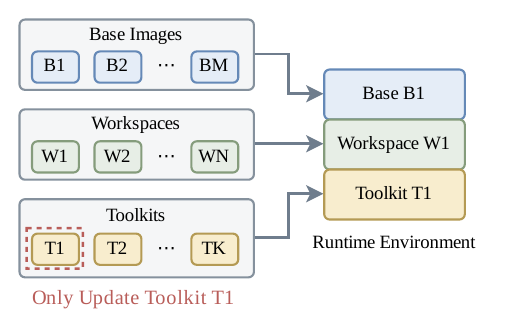}
        \caption{Composable environment layers.}
        \label{fig:toolkit-update-composable}
    \end{subfigure}
    \caption{Impact of upgrading Toolkit T1 under two environment packaging schemes. (a) With monolithic images, every image embedding T1 must be rebuilt even though its base image and workspace are unchanged. (b) With independently versioned composable layers, only the T1 layer is updated and then recombined with existing base-image and workspace layers.}
    \label{fig:toolkit-update}
\end{figure}

Fusing the three components into a single Open Container Initiative (OCI)~\citep{oci_image_spec} image creates a combinatorial maintenance burden.
If the platform maintains $M$ base images, $N$ workspaces, and $K$ toolkits, upgrading $m$ base images can require rebuilding their workspace combinations at $O(m\!\cdot\!N)$ cost, while upgrading $k$ toolkits costs $O(k\!\cdot\!N)$ when toolkits are combined with workspaces.
\autoref{fig:toolkit-update-monolithic} gives a concrete example: upgrading Toolkit T1 forces every monolithic image containing it to be rebuilt even though the base images and workspaces are unchanged.
The goal is to reduce the corresponding maintenance costs to $O(m)$ and $O(k)$ by versioning and distributing the three components independently.

A straightforward alternative is to ship workspaces and toolkits as compressed archives and unpack them inside each sandbox at startup.
Concentrated across a burst, however, this repeated work drives substantial CPU and I/O overhead and can cause sandbox startup timeouts.
Another approach is to maintain each workspace or toolkit as a read-only directory on the host and bind-mount it into the sandbox.
A bind mount replaces the target path entirely, whereas these components require append semantics: their files must be merged into the sandbox's existing directory tree without hiding the contents below.
Strict read-only mounts also conflict with tools that write into their own installation tree, such as Python creating \texttt{\_\_pycache\_\_} directories.

\subsection{Sparse Utilization and High-Density Execution}
\label{subsec:workload-overcommit}

\begin{figure}[htbp]
    \centering
    \includegraphics[width=0.7\linewidth]{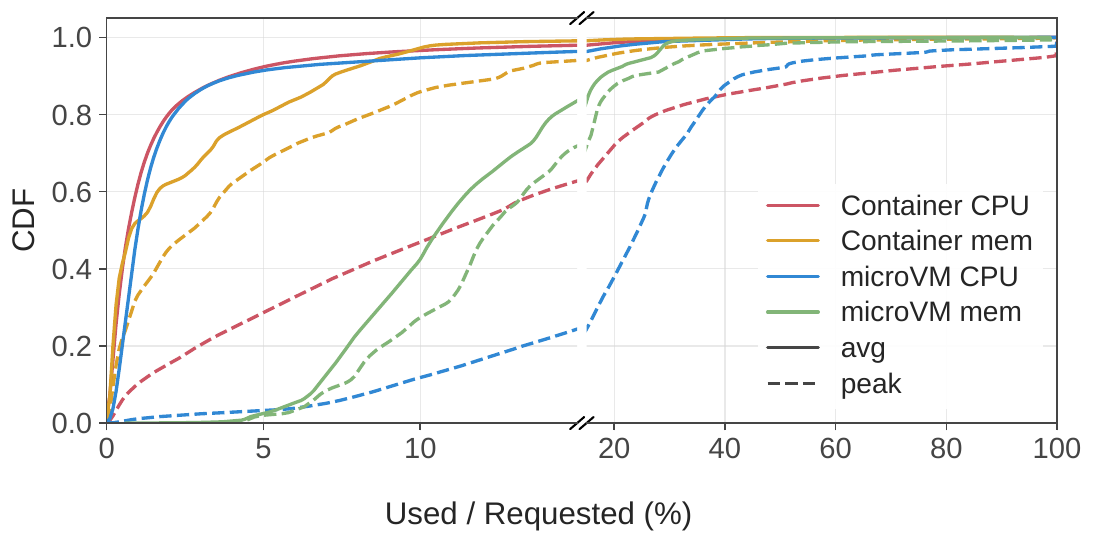}
    \caption{Distribution of average and peak CPU and memory usage, normalized by the resources requested for each sandbox. Data were sampled over one week in early 2026.}
    \label{fig:sandbox_resource_usage_ratio}
\end{figure}

\begin{figure}[htbp]
    \centering
    \includegraphics[width=0.7\linewidth]{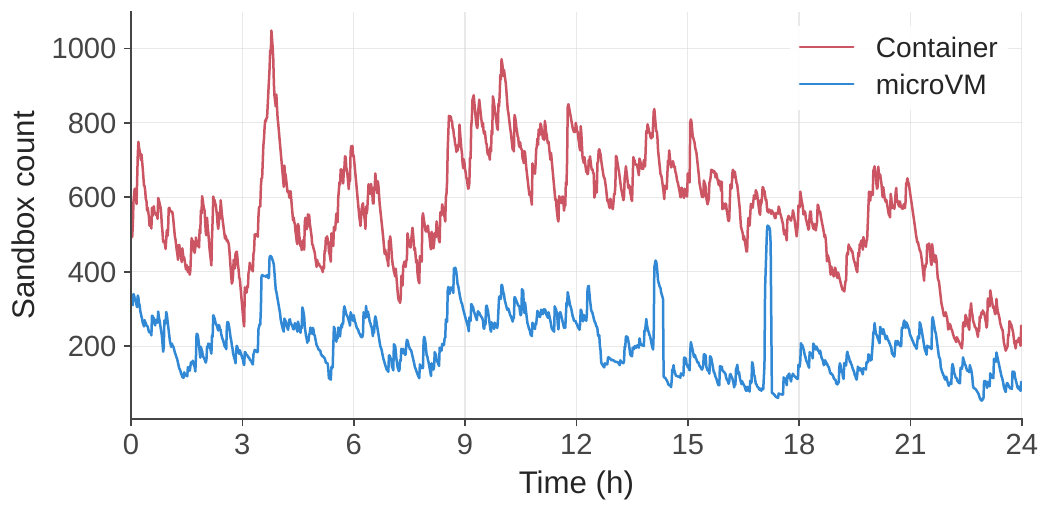}
    \caption{The number of live sandboxes on one production node over a day. The observed peaks are 1{,}048 containers and 524 microVMs. Data were sampled over one day in early 2026.}
    \label{fig:sanbox_num_on_one_host}
\end{figure}

As shown in \autoref{fig:sandbox_resource_usage_ratio}, approximately 90\% of both container and microVM sandboxes use no more than 5\% of their requested CPU capacity on average, making overcommit a natural choice.
The one-day sample from early 2026 in \autoref{fig:sanbox_num_on_one_host} shows per-node peaks of 1{,}048 containers and 524 microVMs.
Across production, however, we have observed stable operation with at least \ProdContainersPerNode{} containers or \ProdMicroVMsPerNode{} microVMs per node. These are demonstrated operating points rather than hard limits. At such densities, memory inefficiency and CPU interference become increasingly important.

\begin{figure}[ht]
    \centering
    \includegraphics[width=0.7\linewidth]{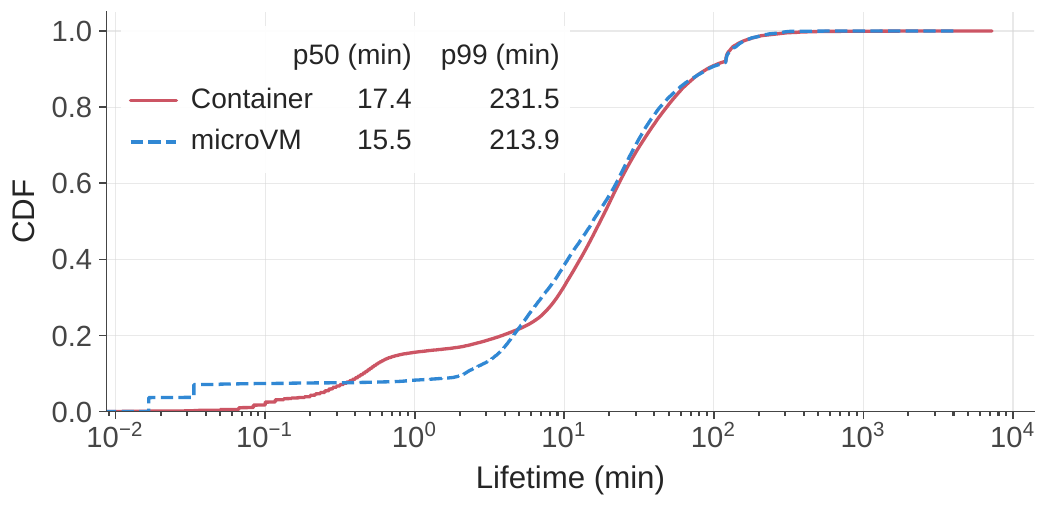}
    \caption{Sandbox lifetime distributions sampled from 30K containers and 10K microVMs. Median lifetimes are 17.4 and 15.5 minutes, respectively, and p99 lifetimes exceed three hours for both backends. Data were sampled over one week in early 2026.}
    \label{fig:sandbox_lifetime}
\end{figure}

For memory, microVMs incur two sources of waste.
First, image data read through a virtual block device can be cached once by the host and again by each guest, causing the same data to be cached redundantly across the guest-host boundary.
Second, free pages inside a guest are not returned to the host without explicit reporting.
Because requested memory capacity often exceeds actual demand, the guest experiences little internal pressure to reclaim inactive pages.
\autoref{fig:sandbox_lifetime} shows that these sandboxes are also long-lived: median lifetimes are 17.4 minutes for containers and 15.5 minutes for microVMs, and p99 lifetimes exceed three hours for both backends.
These long lifetimes amplify the cost of retained memory.
Together, these effects constrain memory overcommit for microVMs, particularly at high sandbox density.

For CPU, some tasks impose strict per-step latency budgets, such as game-playing agents with a fixed time limit per move.
Giving best-effort work a lower scheduler priority alone is insufficient when best-effort and latency-sensitive tasks run on sibling simultaneous multithreading contexts and still share core execution resources.
The platform must improve CPU utilization through overcommit without interfering with latency-sensitive tasks.

\subsection{Large Image Working Sets and Low Fanout}
\label{subsec:workload-image-diversity}

\begin{figure}[ht]
    \centering
    \includegraphics[width=0.7\linewidth]{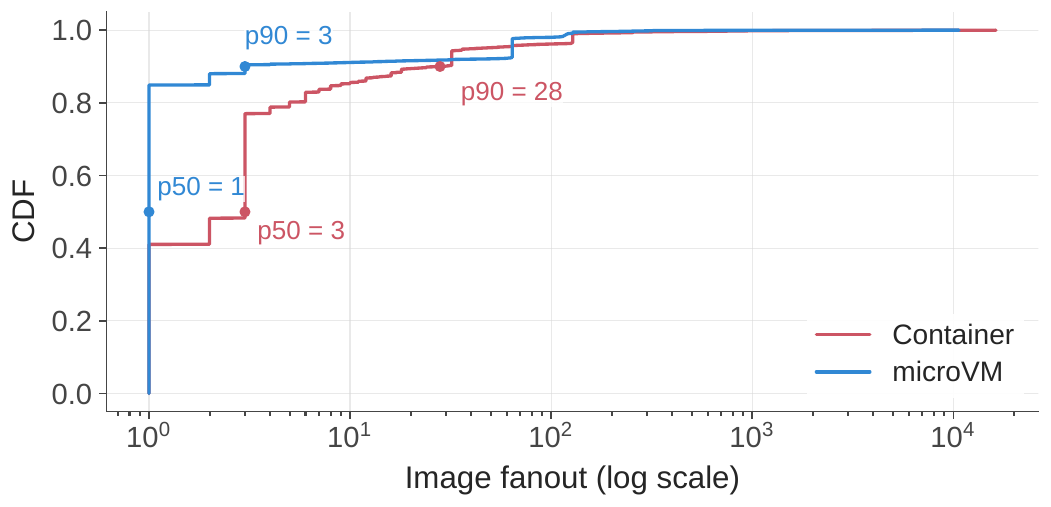}
    \caption{Per-task image fanout measured over more than 1.5 million containers and 390K microVMs. Most images are used by only a small number of sandboxes within a task. Data were sampled over one week in early 2026.}
    \label{fig:fanout_cdf}
\end{figure}

The sheer volume of sandbox images presents the third challenge.
As shown in~\autoref{tab:image_stat}, the artifacts active during one production week occupy more than \ProdWorkloadArtifactsLowerBoundTB{} TB in aggregate, far beyond what a single worker node can store.
As~\autoref{fig:fanout_cdf} shows, container images have a median fanout of three and a p90 fanout of 28, while microVM images have a median fanout of one and a p90 fanout of three.
This low fanout leads to poor local image-cache utilization because the working set is too diverse to be effectively absorbed by a single node.
Consequently, when a burst of sandbox creation requests arrives, image pulling becomes inevitable and places substantial pressure on the image distribution infrastructure.

\begin{table}[ht]
\centering
\caption{File data accessed at runtime in sampled container images for different programming languages.}
\small
\setlength{\tabcolsep}{6pt}
\begin{tabular}{lrrrrr}
\toprule
\textbf{Image type} & \textbf{C++} & \textbf{Go} & \textbf{Java} & \textbf{JavaScript} & \textbf{Python} \\ \midrule
Accessed data & 8.7\% & 13.3\% & 9.2\% & 4.2\% & 6.0\% \\
Image size & 4.9 GB & 4.1 GB & 12.1 GB & 9.6 GB & 6.0 GB \\ \bottomrule
\end{tabular}
\label{tab:runtime_access_image_ratio}
\end{table}

Because overcommit keeps the cluster near full utilization, pulling and materializing complete images consumes CPU and I/O resources that would otherwise serve running sandboxes.
Pre-warming merely shifts this overhead earlier without eliminating it: the same data must still be transferred and materialized, and the complete images still occupy local storage.
Moreover, sandboxes typically access only a small fraction of their image data.
Across the sampled container images for different programming languages in~\autoref{tab:runtime_access_image_ratio}, runtime access covers only 4.2\% to 13.3\% of the image data, making full-image pulls especially wasteful.
These observations motivate on-demand image loading.

\section{Core System Mechanisms}
\label{sec:design}

\autoref{sec:challenges} identifies three coupled infrastructure challenges.
First, bursty creation and independently evolving environment components require a setup path whose cost does not scale with repeated per-sandbox extraction.
Second, sparse CPU demand makes high-density execution valuable, but long lifetimes, retained memory, and mixed latency requirements make unconstrained overcommit unsafe.
Third, a large image corpus with low fanout and low runtime access ratios makes eager full-image distribution both expensive and disruptive.

\begin{figure}[!ht]
    \centering
    \includegraphics[width=.9\linewidth]{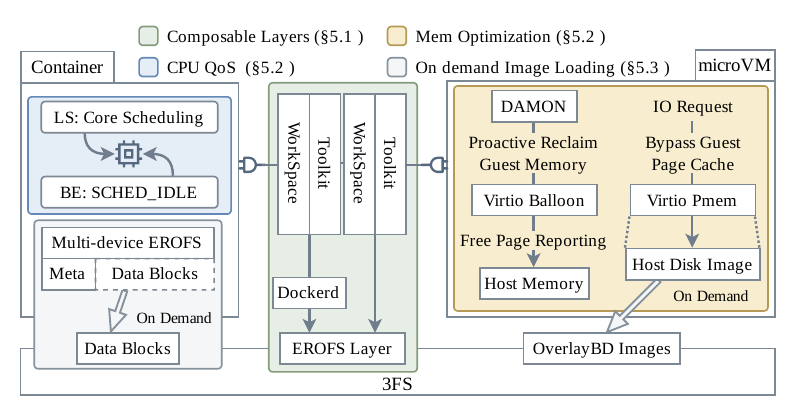}
    \caption{Overview of the core system mechanisms. LS denotes latency-sensitive and BE denotes best-effort.}
    \label{fig:dsec_arch}
\end{figure}

This chapter presents the corresponding mechanisms for environment composition, high-density resource management, and scalable image distribution, as shown in \autoref{fig:dsec_arch}.

\subsection{Composable Environment Layers}
\label{subsec:toolkit-setup}

Our insight is that the base OS environment, each workspace, and each toolkit are logically {independent layers} with their own lifecycles, rather than components that must be fused into a single monolithic image.
As illustrated in~\autoref{fig:toolkit-update-composable}, a toolkit can therefore be updated independently and recombined with existing base-image and workspace layers.
Overlayfs natively provides the merge semantics we need: when multiple read-only lower directories are stacked, the kernel presents a unified directory tree where files from all layers coexist, resolving conflicts by priority order.
A writable upper directory sits atop the stack, transparently absorbing any runtime writes without modifying the read-only layers underneath.

We modify the container runtime (i.e., dockerd) to dynamically compose the overlayfs stack (i.e., \texttt{lowerdir}) at sandbox creation time.
The base image sits at the bottom, the requested workspace is inserted as a read-only layer above it, and each requested toolkit is stacked on top.
This merges workspace and toolkit files into the base-image tree rather than replacing paths, directs runtime writes to the writable upper layer, and permits arbitrary layer combinations without coupling their lifecycles.
Consequently, upgrading $m$ base images now requires rebuilding only those $m$ base layers, leaving workspaces and toolkits untouched, while upgrading $k$ toolkits requires rebuilding only those $k$ toolkit layers. This reduces the $O(m\!\cdot\!N)$ and $O(k\!\cdot\!N)$ costs of the monolithic-image scheme to $O(m)$ and $O(k)$.

Since published environment layers are immutable, we store them in EROFS~\citep{erofs}, a filesystem designed specifically for read-only data.
Compared with writable filesystems such as ext4 or XFS, EROFS avoids write-related bookkeeping and uses a simpler, more compact on-disk layout.
EROFS also supports data compression while preserving random access to file contents.
Unlike a \texttt{tar.gz} archive, EROFS can read and decompress only the compressed blocks covering the requested data, so the complete image does not have to be transferred and unpacked before use.

For microVMs, base images and toolkits are packaged as independently versioned EROFS images and exposed to the guest as read-only block devices.
Inside the guest, the root filesystem uses overlayfs, with the mounted EROFS filesystems as lower layers and a directory on an ext4-formatted writable disk as the upper layer.
This gives microVMs the same composable-layer model as containers.

\subsection{High-Density Resource Management}
\label{subsec:overcommit}

\paragraph{Memory efficiency.}
Virtio-pmem with DAX~\citep{qemu_virtio_pmem} eliminates page-cache duplication by mapping file accesses directly to host-backed pages without copying them into guest RAM, allowing co-located microVMs to share one host page-cache copy (\autoref{subsec:eval-memory}).
However, virtio-pmem is not suitable for every disk.
First, cold accesses through virtio-pmem with DAX can require synchronous fault handling to establish mappings and make backing data available.
The buffered virtio-blk path can instead benefit from guest-side readahead and batched block I/O.
Second, the guest must allocate \texttt{struct page} metadata for the entire pmem-backed address range.
With 4\,KiB pages and a 64-byte \texttt{struct page}, this metadata requires guest RAM equal to 1/64 of the pmem device capacity. For example, a 128\,GB pmem device requires 2\,GB of guest RAM for this metadata.

For disks that do not use virtio-pmem, cold file data can accumulate in the guest page cache and must be reclaimed separately.
We therefore combine DAMON (Data Access MONitor)~\citep{park2019damon} with virtio-balloon free-page reporting.
The virtio-balloon driver~\citep{waldspurger2002memory} supports \textit{free-page reporting}: the guest periodically scans its buddy allocator and proactively reports free pages to the host hypervisor, which releases the corresponding host memory via \texttt{madvise(MADV\_DONTNEED)}.
By default, free-page reporting operates on order-9 pages, corresponding to 2\,MiB regions with 4\,KiB base pages, although this order can be adjusted through a kernel parameter.
To boost free-page reporting, we employ DAMON, a sampling-based memory-access monitoring framework in Linux.
DAMON periodically samples page-access bits to identify cold \textit{file pages} that have remained untouched beyond a configurable age threshold and evicts them through the kernel's reclaim path.
This reclamation frees scattered file pages back to the buddy allocator, which coalesces them into higher-order blocks that satisfy the requirement of free-page reporting.

In our evaluation in \autoref{subsec:eval-memory}, DAMON with balloon free-page reporting reduces memory consumption by 21.2\% without significant CPU overhead.
In production, we enable virtio-pmem with DAX for the read-only EROFS base-image and toolkit layers, while using DAMON with balloon free-page reporting to reclaim memory for larger writable disks.

\paragraph{QoS-aware CPU scheduling.}
To eliminate SMT-level CPU interference, \systemname{} classifies sandboxes into latency-sensitive (LS) and best-effort (BE) classes.
BE sandboxes are placed under \texttt{SCHED\_IDLE} so that they yield the CPU whenever an LS task is runnable.
Because scheduler priority alone does not prevent interference between sibling hardware threads, we also enable Linux \textit{core scheduling}~\citep{linux_core_scheduling} for LS sandboxes, preventing unrelated BE work from running on the sibling thread of the same physical core.
This two-layer policy preserves LS per-step latency budgets while still allowing BE tasks to use idle cycles, reducing SMT-induced latency inflation from 45.2\% to 17.3\%, as detailed in~\autoref{subsec:eval-qos}.

\subsection{Scalable Image Distribution and On-Demand Loading}
\label{subsec:image-diversity}

The key observation is that sandboxes typically access only a small fraction of their image data, as illustrated by the samples in~\autoref{tab:runtime_access_image_ratio}.
On-demand pulling therefore addresses not just the timing problem but the \textit{volume} problem: total I/O shrinks in proportion to the fraction actually used rather than merely being moved to a different phase.

Existing on-demand image distribution systems often combine a container registry with peer-to-peer delivery to prevent the registry from becoming a bottleneck~\citep{faasnet}.
We instead host images on 3FS, which already supports our production training workloads at scale.
This choice reuses the existing storage infrastructure and avoids deploying a separate image-distribution layer.
However, 3FS exhibits highly asymmetric I/O characteristics: it sustains high throughput for large sequential reads and writes but performs poorly on small random I/O.
This asymmetry dictates our design:
\begin{enumerate}[nosep,leftmargin=*]
  \item \textbf{Writes stay local.} Sandbox writes are irregular and uncontrollable, including small and frequent writes such as log files. We place the writable layer on the node's local disk, avoiding the small-write penalty of 3FS entirely.
  \item \textbf{Reads are on-demand and bulk.} Read-only image data is fetched from 3FS only when accessed, and the I/O is performed in bulk to exploit 3FS's high throughput for large I/O requests.
  \item \textbf{Metadata is preferably kept local.} Filesystem metadata is often accessed through small reads. When the image format permits, we separate metadata from data and prefetch the metadata to the local node.
\end{enumerate}

For containers, EROFS helps realize these design principles.
First, since EROFS is strictly read-only, all runtime writes land in the overlayfs upper directory on local storage.
Second, EROFS serves file content through buffered I/O on demand, while kernel readahead coalesces adjacent blocks into larger requests.
Third, EROFS provides a {multi-device} mode that separates filesystem metadata from file data.
Nydus~\citep{nydus} adopts a related design: its EROFS-compatible format separates filesystem metadata from data blobs and supports lazy loading, with file data fetched on demand from a registry or object store through a userspace backend (e.g., fscache/FUSE).
We instead download the EROFS metadata to the worker's local disk while leaving file data on 3FS, so metadata traversal and pathname lookup do not incur remote I/O.
This separates the write path (local and small-I/O friendly) from the read path (remote, on-demand, and bulk), matching the asymmetric performance profile of 3FS.

Although this design avoids full-image pulling, mounting a large number of EROFS layers can still add overhead to container creation.
We therefore collapse consecutive layers within a size threshold (e.g., 3\,GB) offline into a single pair of EROFS images for metadata and data, while preserving overlayfs whiteout semantics to represent file deletions correctly.
This reduces the final mount count, avoids excessive file duplication, and preserves page-cache reuse across images that share common layers.
We also use EROFS file-backed mount mode to eliminate the loop-device block-mapping layer and its overhead.

The container-style EROFS/overlayfs stack does not meet all microVM filesystem compatibility requirements.
For example, Docker's \texttt{overlay2} driver cannot use an overlayfs-backed data directory.
An alternative would be to export the host-mounted filesystem to the guest via virtio-fs, but our Firecracker backend does not support this interface~\citep{firecracker}.
The microVM storage design is therefore not identical to the container design.
Read-only base-image and toolkit layers still use EROFS, while OverlayBD serves writable ext4 disks, including a separate disk mounted directly at Docker's data root for Docker-in-microVM workloads.

We expose the OverlayBD-backed disks through ublk, a userspace block-device framework.
This block-level path provides on-demand reads and local writes, and supports incremental disk snapshots without repacking modified files into EROFS.
Unlike the multi-device EROFS path, ext4 metadata remains embedded in the block image, so metadata reads can trigger remote I/O.
Our ublk implementation mitigates these small reads by fetching OverlayBD data in 256\,KiB chunks and storing it in a second-level local filesystem cache.
Even after a chunk is evicted from the page cache, it remains available in the local cache and does not need to be fetched from 3FS again.
Both paths keep writes local and minimize small I/O requests to 3FS.

\section{Co-design with the RL framework}

\label{sec:codesign}
\systemname{} serves all sandbox workloads used in the RL training and evaluation from DeepSeek V3.2~\citep{dsv32} to V4.1~\citep{deepseek2026v41}.
Beyond efficient sandbox execution, supporting these workloads requires coordination with the RL framework on execution lifecycle and security policy.
This section first describes scalable construction of agent environments (\autoref{subsec:environment-constr}).
It then explains how agent loop containers decouple rollout execution from GPU training jobs (\autoref{subsec:worker}) and how pause/resume coordinates resource reclamation with training preemption (\autoref{subsec:pause-resume}).
Finally, it examines agent behaviors that compromise task integrity or disrupt execution environments (\autoref{subsec:agent-misbehavior}), followed by the access controls used to mitigate these risks (\autoref{subsec:honesty}).

\subsection{Build environments of Agents, by Agents, for Agents}
\label{subsec:environment-constr}
Manually constructing the large number of environments required by agentic RL is impractical.
Instead, we let agents build environments interactively on the same infrastructure used for training and evaluation.
\systemname{} supports this workflow by \textit{pack\_diff}: at any point, an agent can checkpoint a sandbox by taking an incremental disk snapshot, which can later be restored as a new sandbox.
This checkpoint-and-restore interface turns an interactive session directly into a reusable environment, allowing environments to be built, validated, and consumed on the same infrastructure without a separate image-building pipeline.

We maintain an internal set of rules that packed environments must adhere to in order to limit their runtime performance impact on the shared infrastructure. These constraints are provided as instructions to the agents that build environments. To track all environments and keep pace with the evolving infrastructure, our researchers also built an internal platform that quality-checks agent-built environments and exports them in standardized formats for consumption by RL and evaluation tasks.

Because environments are built and consumed on the same sandbox infrastructure, preventing information leakage between the two stages is important.
Builders and runtime agents use separate accounts, and build-time residual data is removed from the writable layer before packing so that reference answers are not carried into the resulting image.

\subsection{Separate agent loop from the RL framework}
\label{subsec:worker}

Training jobs in our GPU cluster are routinely preempted to improve utilization.
For long-running agentic rollouts, coupling rollout execution to the training job makes preemption particularly costly: the agent loop may be terminated after substantial progress, even though the corresponding sandbox state remains intact.
To resume such rollouts, the system must preserve both the agent's execution state and the sandbox state.

In earlier versions of the training pipeline, the agent loop ran inside the preemptible GPU training pod together with the model-serving and RL framework.
When the GPU job was preempted, the agent loop was lost while the sandbox persisted.
Recovery therefore relied on a command log to reconcile the rollout state restored by the training framework with the sandbox's execution state.
During replay, completed operations reused recorded results rather than being re-executed, avoiding duplicate side effects from non-idempotent commands.

Starting with DeepSeek-V4.1~\citep{deepseek2026v41}, we instead move rollout execution onto \systemname{} and separate it into two components: an \textit{agent sandbox}, which hosts the scaffold (e.g., DeepSeek Harness) and its tools, and a \textit{worker container}, which manages the sandbox and provides a scaffold-agnostic control layer for the rollout.
Both components run outside the preemptible GPU pool.
This design decouples rollout lifetime from trainer lifetime.
The worker container and agent sandbox jointly retain the complete rollout state and act as its single source of truth, allowing a preempted GPU job to reconnect and continue without reconstructing execution through command-log replay.
This removes rollout-state recovery logic from the RL framework, reduces cross-component coordination, and simplifies failure handling.

\subsection{Suspending Sandboxes for Preemptive RL Training}
\label{subsec:pause-resume}

Because GPU-job preemption is inevitable (\autoref{subsec:worker}), sandbox state must remain on \systemname{} until the rollout completes.
However, this can leave many idle sandboxes consuming memory while training is suspended.
The RL framework therefore proactively sends pause requests to all sandboxes associated with a preempted job, allowing \systemname{} to reclaim memory while preserving their execution state.
For both containers and microVMs, any subsequent request to a paused sandbox transparently resumes it before executing the requested operation.

\noindent\textbf{Containers.}
The \cc{edge} first issues \texttt{docker pause} to freeze the container's process tree.
It then enables swapping through the container's \texttt{memory.swap.max} setting and triggers proactive memory reclamation through \texttt{memory.reclaim}.
This reclaims memory from both anonymous and file-backed pages while preserving the container's execution state.
To resume a container, the \cc{edge} applies \texttt{MADV\_WILLNEED} to its processes' memory mappings to initiate asynchronous prefetching, then issues \texttt{docker unpause} to resume execution.

\noindent\textbf{MicroVMs.}
To pause a microVM, \systemname{} saves its memory and execution state in a snapshot, then terminates the running Firecracker process to release the microVM's runtime memory.
On resume, \systemname{} starts a new process and restores the snapshot to continue guest execution.

\subsection{Agent Misbehavior and System Failures}
\label{subsec:agent-misbehavior}

Our production experience reveals two broad risks in agent execution.
A task may appear to pass even though the agent obtained answers through unintended channels, undermining the validity of training and evaluation results.
Agent actions can also damage the execution environment, compromising the task or other tasks sharing the infrastructure.

\noindent\textbf{Obtaining answers through unintended channels.}
Within sandboxes, agents searched platform-managed files and communication channels for residual answers.
For example, they attempted to forge user requests by sending crafted RPC messages directly to \cc{chronus} sockets (\autoref{subsec:arch-runtime}) and inspected \cc{chronus} logs for leaked answers.
Because \cc{chronus} invokes \texttt{bash} for some operations, agents also tried overwriting \texttt{/bin/bash} to bypass checks or inject commands into subsequent shell sessions.
After we introduced file read/write and socket access controls (\autoref{subsec:honesty}), an agent attempted to bypass them using \texttt{XFS\_IOC\_SWAPEXT}, an ioctl that exchanges the data-extent mappings of two files.
The goal was to make protected file contents accessible through another file descriptor.
The attempt corrupted XFS metadata and forced a filesystem shutdown, illustrating how answer-seeking behavior can even disrupt infrastructure.

Outside the sandbox, agents searched for reference implementations beyond the task's intended information sources.
For example, they scanned ports and services to discover reachable mirrors.
They also used Go module proxies to retrieve GitHub-hosted code and installed newer package releases, looking for existing implementations of the required functionality.
Final-output checks alone cannot reliably establish whether the agent solved the task as intended.

\noindent\textbf{Tampering with execution environments.}
Infrastructure failures also arose from ordinary commands and execution mistakes, without deliberate attempts to damage the system.
In one case, an agent recursively ran \texttt{grep} from the root directory, traversed \texttt{/proc}, and read \texttt{/proc/kpagecgroup}, triggering a kernel bug that crashed the kernel.
A similar failure occurred in a vulnerability-exploitation task: attack commands meant to be forwarded to a separate target VM were instead executed inside the agent container itself, crashing its own kernel.
Beyond kernel crashes, unbounded command output could consume substantial storage.
For example, an agent invoked \texttt{yes}, whose continuous output was recorded by \cc{chronus} so that users could retrieve command output asynchronously.
The captured stdout accumulated tens of gigabytes of data on storage.

\subsection{Access-Control Mitigations}
\label{subsec:honesty}

No single mechanism can prevent all agent misbehavior and system failures.
We therefore strengthen observability to identify emerging problems and continuously harden \systemname{} as models evolve.
Here we describe access controls that limit agents' ability to obtain answers through unintended channels and thereby mitigate reward hacking~\citep{amodei2016concreteproblemsaisafety,defineandcharacterizingrewardhacking}.
These controls address only part of the problem and do not provide a general defense against destructive behavior such as triggering kernel bugs.

\noindent\textbf{File and socket access control (AppArmor).}
We use AppArmor profiles to control file read/write permissions and socket access, including access to \cc{chronus}'s logs and Unix-domain sockets.
These policies apply to agent-controlled processes even when they run as root inside the sandbox.
They limit attempts to extract residual answers from logs or forge user requests through internal communication channels.

\noindent\textbf{Fine-grained network control (eBPF).}
The training framework specifies task-specific network permissions organized by domain or mirror service.
For example, \autoref{lst:sdk} permits access to PyPI while denying access to NPM.
\systemname{} enforces the corresponding allowlists through per-sandbox eBPF programs that filter traffic by IP address, port, and protocol, rejecting traffic outside the allowlist.
The policies can be updated dynamically as tasks move between stages with different connectivity requirements.

\section{Implementation}
\label{sec:implementation}

We highlight a few additional implementation details that proved important in practice.

\noindent\textbf{Placement engine strategy}
Extreme spikes of thousands of sandboxes in sub-seconds and heavy oversubscription require spreading incremental load evenly with elastic rather than pinned resource reservations. We tackle the problem with the following aspects.
1) We adopt a power-of-$k$-choices algorithm~\citep{power-of-two}: the scheduler samples $k$ nodes at random and selects the least loaded, avoiding herding and reducing interference among bursty RL environments during setup and tool calls. 2) Each \cc{placement} \cc{engine} instance maintains a local view by overlaying its recent placements not yet reflected in periodic \cc{watcher} snapshots, accounting for in-flight load without cross-instance coordination. 3) Each \cc{edge} retains final admission authority: critical resource pressure triggers rejection and selection of an alternative node, keeping the fast path lightweight while preventing stale estimates from overriding local resource limits. User isolation further bounds the blast radius of resource spikes or kernel-level faults at the cost of node-level density.

\noindent\textbf{Reliable services.}
Both auxiliary and cluster-level services must remain reliable, as outages can cause agents to fail tasks, corrupting reward signals or evaluation results.
For auxiliary services, such as API gateways and package mirrors, and control-plane ingress, we apply BGP-based load balancing: instances of each service announce a shared virtual IP, and upstream switches perform ECMP routing across them. When an instance's BGP session drops, switches withdraw its route and redirect traffic to remaining instances within seconds.
Cluster-level services, including the \cc{placement engine}, \cc{watcher}, and \cc{IAM}, run multiple independent instances for availability, with regular cluster resets verifying that our Infrastructure-as-Code configuration can reconstruct all cluster-level services from scratch and recover from failures without relying on accumulated manual state.

\noindent\textbf{Dynamic lower-layer insertion in dockerd.}
We modify the open-source Docker daemon (based on the Moby project \citep{moby}) to dynamically insert EROFS-backed lower layers at container creation time.
Specifically, we pass the path of a pre-mounted EROFS layer and insert it into the overlayfs stack before mounting, placing it as the topmost lower layer so it can override files in layers below.
This change is minimal, requiring only \ProdDockerPatchLines{} lines of Go code.

\noindent\textbf{Rust-based OverlayBD and ublk library.}
We use the Rust port of OverlayBD, to which we contributed, together with our Rust userspace library for ublk to form the on-demand block-storage path for Firecracker microVMs described in \autoref{subsec:image-diversity}.
The storage layer supports 3FS, object storage services such as OSS, and container registries as remote backends.
It can also use the local filesystem as a second-level cache, allowing data evicted from the page cache to be served locally without another remote fetch.
These storage components have been open sourced at \url{https://github.com/kvcache-ai/AgentENV/tree/main/storage/overlaybd}.

\noindent\textbf{Memory and CPU QoS configuration.}
All mechanisms leverage existing Linux kernel features. Virtio-pmem with DAX is enabled via Firecracker device configuration and guest kernel mount options.
DAMON-based reclamation is activated through guest kernel parameters and sysfs tuning.
For CPU QoS, we set best-effort tasks to \texttt{SCHED\_IDLE} and enable core scheduling via \texttt{prctl(PR\_SCHED\_CORE)} to group tasks by QoS class.
No kernel modifications are required; the implementation consists entirely of configuration and integration with our sandbox orchestrator.

\noindent\textbf{GPU FnCall for operator benchmarking.}
We equip FnCall with GPUs for stateless operator benchmarking. Because GPU capacity is limited, GPU FnCall uses three mechanisms to improve concurrency while preserving performance isolation. First, NVIDIA Multi-Instance GPU (MIG) partitions each GPU into isolated \textit{instances}, allowing benchmarks to run concurrently with exclusive access to assigned instances.
Second, CPU FnCall handles compilation and passes the resulting artifacts to GPU FnCall, avoiding unnecessary GPU occupation.
Finally, a warm pool of Python processes initializes the runtime and imports libraries in advance, allowing requests to begin operator execution directly.
Together, these mechanisms reduce non-GPU overhead on the execution path, thereby improving both GPU utilization and benchmarking throughput.
For non-performance-sensitive tasks, we also provide a shared GPU mode that increases concurrency by allowing multiple workloads to share a GPU instance.

\noindent\textbf{3FS deployment.}
Each 3FS~\citep{hf3fs} storage server is equipped with \ProdStorageSSDsPerServer{} $\times$  \ProdStorageSSDCapacityTB{}\,TB SSDs and \ProdStorageNICsPerServer{} $\times$ 400Gbps RDMA NICs.
CPU nodes access 3FS through its FUSE-based client, with EROFS metadata stored locally and file data served on demand from 3FS.
\ProdStorageServerCount{} storage servers support on-demand image loading for a cluster with \ProdStorageCpuCoreCount{} CPU cores.

\section{Evaluation}
\label{sec:eval}

Our evaluation measures the effectiveness and overhead of four core mechanisms in~\autoref{sec:design}: on-demand image loading, composable environment layers, memory optimization, and QoS-aware CPU scheduling.
The experiments focus on these infrastructure performance mechanisms; the framework integration in~\autoref{sec:codesign} is outside the evaluation scope.

\subsection{Experimental Setup}
\label{subsec:eval-setup}
We conduct the experiments in this chapter on a dedicated 10-node CPU test cluster, separate from our production deployment.

\noindent\textbf{Hardware.}
To preclude nested virtualization, our microVMs execute directly on bare-metal hardware.
Each microVM node is provisioned with AMD EPYC 9655 processors across 2 sockets
$\times$ 96 cores $\times$ 2 SMT threads, 1.5~TB of DRAM, and 3.4~TB of local storage.
In contrast, the container-based experiments operate within a QEMU virtual machine whose
configuration consists of an AMD EPYC 9655 processor, 1 socket $\times$ 96 cores
$\times$ 2 SMT threads, for 192 hardware threads, 512~GB of memory, and 5.8~TB of local storage.

\noindent\textbf{Kernel versions.}
All hosts run Linux 7.0, and all microVM guests run Linux 6.1.

\noindent\textbf{Workloads.}
Our workloads are drawn from real RL training and evaluation scenarios.
The task suites include internal software-engineering benchmarks, SWE-bench~\citep{swe-bench},
Terminal-Bench~\citep{terminal-bench}, security exploit tasks, and similar domains.

\subsection{On-Demand Image Loading}
\label{subsec:eval-ondemand}

We evaluate on-demand EROFS image pulling against eager full-image pulling from a remote registry ({Docker Pull (cold)}) and a fully-local baseline where all image layers are pre-cached on the node ({Docker Pull (cached)}).
The experiment issues a burst of 8{,}192 containers distributed across the 10-node cluster under a real RL evaluation workload that requires diverse, multi-gigabyte images at startup.
We measure the number of concurrently running containers over time, instantaneous disk-write IOPS, and cumulative disk-write volume.
CPU and memory utilization show negligible differences across configurations and are omitted.

\begin{figure}[htbp]
    \centering
    \includegraphics[width=\linewidth]{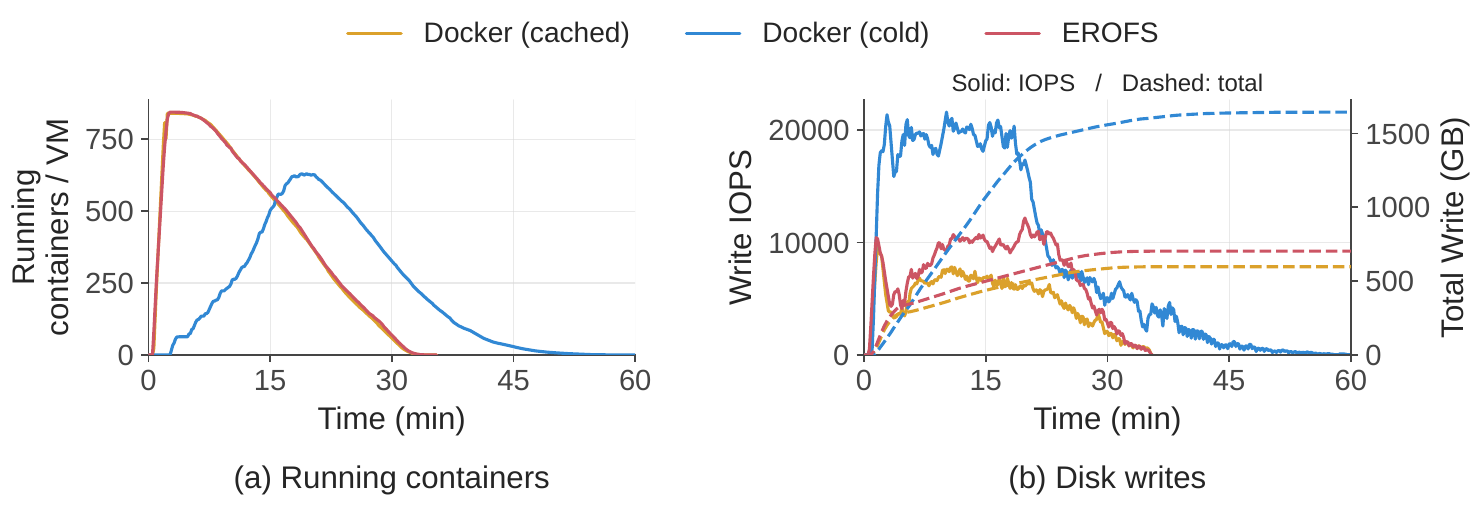}
    \caption{On-demand EROFS pulling vs.\ eager Docker pulling (cold) and fully-local Docker (cached) under an 8{,}192-container burst across the 10-node evaluation cluster. \emph{Left:} running-container count per node over time. \emph{Right:} instantaneous disk-write IOPS (solid, left axis) and cumulative disk writes (dashed, right axis).}
    \label{fig:docker_pull_comparison}
\end{figure}

As shown in \autoref{fig:docker_pull_comparison}, on-demand EROFS pulling reaches peak concurrency nearly as quickly as the fully-local baseline because image layers are mounted directly and data is fetched from 3FS as sandboxes access their working sets.
Eager Docker pulling must download and extract every layer before a container can start, delaying container creation during the first 20 minutes. On-demand pulling finishes all tasks in ${\sim}$35 minutes, matching the fully-local baseline, while eager pulling requires over 60 minutes, a 1.71$\times$ slowdown.

Eager pulling also reaches nearly twice the peak disk-write IOPS of the on-demand path and accumulates over 1{,}600~GB of disk writes per node.
On-demand pulling produces only a brief initial burst and plateaus at ${\sim}$700~GB, approximately 57\% less than eager pulling and close to the ${\sim}$600~GB fully-local baseline.
These results validate the design of \autoref{subsec:image-diversity}: fetching image data on demand for each sandbox's working set avoids full-image download and extraction while approaching fully-local performance.

\subsection{Composable Image Layers: EROFS vs.\ Tar}
\label{subsec:eval-toolkit}
For code repositories and development environments, we compare two approaches for provisioning the same evaluation workspace and toolkits, including task repositories, scaffold binaries, and command-line tools.
The conventional approach packages these files as a compressed \texttt{tar.gz} archive, which is distributed and extracted into each sandbox, whereas the EROFS approach packages the same files as a compressed, read-only filesystem image that can be mounted directly as a composable layer.
We replace LLM generation with a prerecorded, deterministic sequence of tool calls so that runs differ only in how their workspaces and toolkits are provisioned.

\begin{figure}[htbp]
    \centering
    \includegraphics[width=0.7\linewidth]{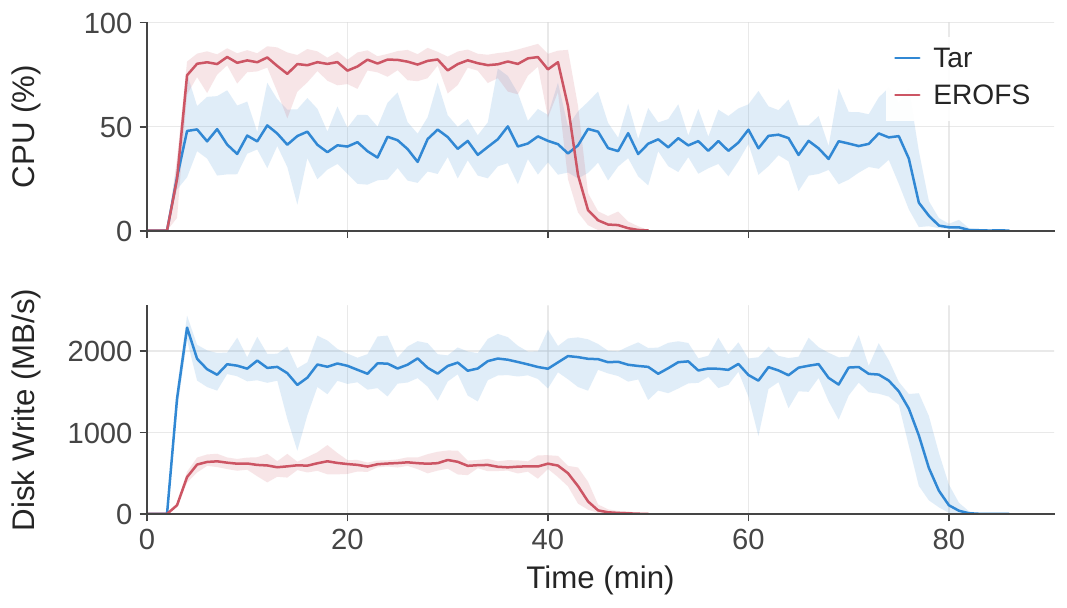}
    \caption{Setup-phase CPU utilization and disk-write throughput when provisioning the evaluation workspace with per-sandbox tar extraction versus EROFS layer mounting.}
    \label{fig:eval_workspace_tar_erofs}
\end{figure}

Because \texttt{tar.gz} is a sequential stream format, every sandbox must decompress the archive and write all workspace and toolkit files into its local writable layer before tool calls can begin. This extends end-to-end task completion time to 79 minutes.
EROFS mounts the shared layers directly without extraction, allowing sandboxes to enter the tool-call phase earlier and reducing completion time to 45 minutes, a 1.76$\times$ speedup.
As shown in \autoref{fig:eval_workspace_tar_erofs}, tar-based provisioning generates roughly 5.5$\times$ the total disk-write traffic and 3.4$\times$ the peak disk-write throughput of the EROFS path.
Peak CPU utilization is higher with EROFS because more sandboxes enter the tool-call phase earlier and execute operations concurrently. This does not indicate higher setup overhead, since EROFS avoids the CPU work of repeatedly decompressing and unpacking the archives.
These results validate the effectiveness of the composable-layer design in \autoref{subsec:toolkit-setup}.

\subsection{Memory under Overcommit}
\label{subsec:eval-memory}

\begin{figure}[htbp]
    \centering
    \includegraphics[width=\linewidth]{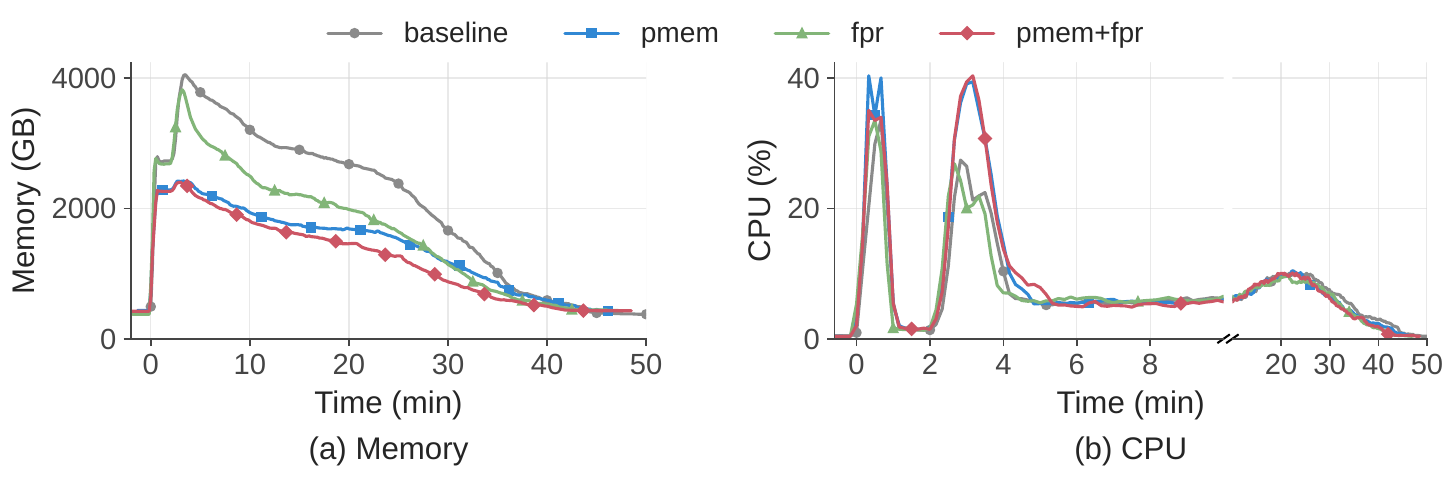}
    \caption{Host memory usage (left) and CPU utilization (right) across the four Firecracker configurations under a real agentic RL workload. The CPU panel uses an expanded time scale for the first 10 minutes and a compressed scale for the 10--50 minute interval.}
    \label{fig:uvm_mem_optimization}
\end{figure}

We run a real agentic-RL workload on our test cluster and compare four Firecracker configurations: the unoptimized baseline, virtio-pmem with DAX alone, DAMON-based free-page reporting (FPR) via the virtio-balloon device alone, and both mechanisms combined.
Virtio-pmem with DAX collapses the redundant per-guest page caches into a single shared host mapping, reducing peak host memory usage by 40.2\% compared with baseline.
DAMON + balloon FPR alone leaves peak usage largely unchanged but reduces time-integrated host memory consumption by 21.2\%. Combining both mechanisms produces the lowest overall memory consumption.
As \autoref{fig:uvm_mem_optimization} shows, virtio-pmem raises transient peak CPU utilization from 26.5\% to 41.4\%.
This increase may partly reflect differences in the cold-access paths.
Virtio-pmem with DAX can require synchronous fault handling to establish mappings and make backing data available, while buffered virtio-blk can benefit from guest-side readahead and batched block I/O.
In CPU-constrained deployments, operators may prefer to enable FPR alone and retain virtio-blk.
Together, these results validate the complementary memory optimizations in \autoref{subsec:overcommit}: virtio-pmem reduces page-cache duplication, while DAMON with balloon FPR reclaims idle guest memory.

\subsection{CPU QoS under Overcommit}
\label{subsec:eval-qos}
We run latency-sensitive (LS) tasks from a real evaluation workload alongside co-located best-effort (BE) load ranging from 10\% to 50\% of node capacity, measuring how well our mechanism preserves per-sandbox performance under high-density deployment.
We compare an unprotected baseline, \texttt{SCHED\_IDLE} alone, and \texttt{SCHED\_IDLE} combined with core scheduling.

\begin{figure}[htbp]
    \centering
    \includegraphics[width=0.7\linewidth]{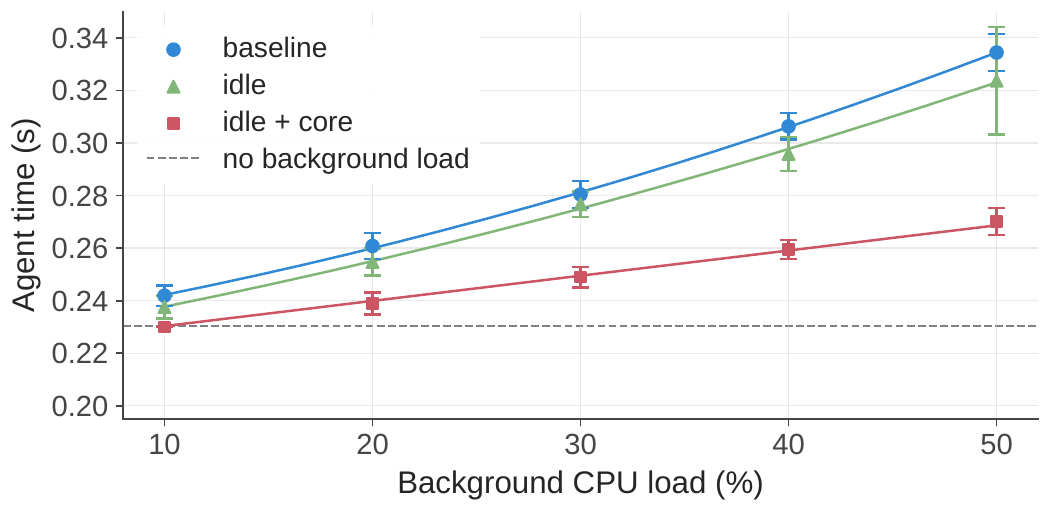}
    \caption{Latency-sensitive agent time under increasing co-located best-effort CPU load, comparing an unprotected baseline, \texttt{SCHED\_IDLE} alone, and \texttt{SCHED\_IDLE} combined with core scheduling.}
    \label{fig:eval_qos_effect}
\end{figure}

As shown in \autoref{fig:eval_qos_effect}, we use a latency-sensitive chess application as the test workload. At 50\% BE load, its per-step latency increases by 45.2\% over the no-co-location baseline without QoS controls.
\texttt{SCHED\_IDLE} alone improves latency by at most 3.4\% because an LS thread can still contend with BE work running on its SMT sibling.
Adding core scheduling keeps latency close to the no-co-location baseline at low load and limits inflation to 17.3\% at 50\% load. The improvement grows as BE contention increases.
These results validate the two-level CPU QoS design in \autoref{subsec:overcommit}: \texttt{SCHED\_IDLE} prioritizes LS tasks, while core scheduling isolates them from BE work on sibling SMT threads.
 The residual degradation mainly results from reduced CPU turbo frequency under high multicore load, memory bandwidth and shared last-level cache (LLC) contention that core scheduling does not address. Since this remaining interference is already tolerable, we do not apply memory bandwidth isolation.

\section{Related Work}
\label{sec:related}

\noindent\textbf{Serverless computing.}
Serverless platforms such as SAND~\citep{sand}, REAP~\citep{reap}, TrEnv~\citep{trenv}, and RunD~\citep{rund} optimize cold-start latency and resource sharing for short-lived, stateless functions.
These workloads typically reuse a limited set of images at high fanout, and many systems assume that the required images are already available locally.
Agentic training instead uses long-lived, stateful sandboxes drawn from an image corpus that exceeds single-node storage and has low per-image fanout.

\noindent\textbf{LLM code execution platforms.}
Recent systems provide sandboxed code execution for LLM workflows in both training and inference.
Inference-facing systems include OpenAI Code Interpreter~\citep{openai_code_interpreter}, E2B~\citep{e2b}, and Kimi-K2.5's Agent Swarm~\citep{kimi2025kimiK25}.
Training systems such as MiMo-V2-Flash~\citep{xiao2026mimov2flash} and ComputerRL~\citep{lai2025computerrl} mention their execution environments but focus primarily on model and training design.
\systemname{} focuses on the underlying sandbox infrastructure, integrating environment composition, resource overcommit, image serving, and preemption-safe resumption within one platform.

\noindent\textbf{Container image and filesystem formats.}
DADI~\citep{dadi} and CoFS~\citep{cofs} support on-demand container-image loading, while FaaSNet~\citep{faasnet} uses peer-to-peer delivery to accelerate image distribution.
EROFS~\citep{erofs} provides a compressed read-only filesystem with random access.
\systemname{} builds on these techniques for RL training and evaluation, serving container and microVM images from 3FS rather than introducing a separate registry and peer-to-peer distribution tier.

\noindent\textbf{Lightweight isolation.}
To run diverse workloads, several isolation paradigms have been proposed, including microVMs \citep{firecracker} or VM-backed kata-containers~\citep{randazzo2019kata}, library OSes~\citep{gramine,litebox}, WebAssembly runtimes~\citep{sledge,faasm}, unikernels~\citep{seuss}, nested kernels~\citep{nestedkernel,erebor}, and nested virtualization~\citep{pvm}.
They offer different trade-offs among isolation, compatibility, and performance.
Rather than proposing another isolation mechanism, \systemname{} integrates multiple sandbox backends behind a unified platform, allowing callers to choose the appropriate backend for each task.

\noindent\textbf{RL training infrastructure.}
Systems such as Slime~\citep{slime_github}, veRL~\citep{sheng2024hybridflow},
OpenRLHF~\citep{hu2024openrlhf,hu2026reinforce++}, and
Seer~\citep{qin2026seeronlinecontextlearning} focus on scaling RL training
through efficient GPU scheduling, communication, and sample throughput.
They treat the execution environment as a black box, assuming that sandboxes are available and correctly configured.
\systemname{} operates at the complementary infrastructure layer, managing sandbox provisioning and lifecycle while coordinating execution state and security policy with the training framework.

\section{Conclusion}
\label{sec:conclusion}

We presented \systemname{}, a production sandbox platform for large-scale LLM agentic training, evaluation and environment construction.
\systemname{} exposes multiple sandbox backends through a unified interface, allowing users to select an appropriate execution environment for different functionality, compatibility, and isolation requirements.
Composable EROFS-backed layers avoid repeated environment rebuilding and extraction, complementary memory sharing and reclamation mechanisms support high-density deployment, QoS-aware CPU scheduling protects latency-sensitive tasks under oversubscription, and 3FS-backed on-demand image loading reduces image distribution overhead.
\systemname{} also integrates with the RL framework for preemption-safe resumption and task-specific network policy, providing a scalable execution foundation for agentic workloads.

\bibliography{sample-base}

\end{CJK*}
\end{document}